\documentclass[10pt,journal,compsoc]{IEEEtran}
\ifCLASSOPTIONcompsoc
  \usepackage[nocompress]{cite}
\else
  \usepackage{cite}
\fi

\usepackage{amsmath,amssymb,amsfonts}
\usepackage{graphicx}
\usepackage{textcomp}
\usepackage{booktabs}
\usepackage[dvips, bookmarks, colorlinks=true]{hyperref}

\usepackage{indentfirst}
\usepackage[noend]{algorithmic}
\usepackage[ruled, vlined, linesnumbered]{algorithm2e}
\usepackage{graphics}
\usepackage{epsfig}
\usepackage{subfigure}
\graphicspath{{fig/}}
\usepackage{comment}
\usepackage{multicol}
\usepackage{tipa}
\usepackage{afterpage}

\usepackage[usenames, dvipsnames]{color}
\usepackage{colortbl}

\newcommand{\name}{\texttt{PiggyBackup}}
\ifx\isthesis\undefined
        \newcommand{\secc}{\section}
        \newcommand{\sect}{Section}
        \newcommand{\sectlower}{section}
        \newcommand{\article}{paper}
        \newcommand{\subsecc}{\subsection}
        
\else
        \newcommand{\secc}{\chapter}
        \newcommand{\sect}{Chapter}
        \newcommand{\sectlower}{chapter}
        \newcommand{\article}{thesis}
        \newcommand{\subsecc}{\section}
        
\fi

\begin{document}
\title{Efficient Network Function Backup by Update Piggybacking}
\author{Kate~Ching-Ju~Lin,~\IEEEmembership{Senior Member,~IEEE,}
        Ruei-Yong~Hong
        and~Yu-Chee~Tseng,~\IEEEmembership{Fellow,~IEEE}% <-this % stops a space
\IEEEcompsocitemizethanks{\IEEEcompsocthanksitem K. C.J. Lin is
with the Department of Computer Science, National Chiao Tung
University, HsinChu, Taiwan, 300.\protect\\
E-mail: katelin@cs.nctu.edu.tw
\IEEEcompsocthanksitem R.-Y. Hong and Y.-C. Tseng are with National
Chiao Tung University.}% <-this % stops an unwanted space
}

%\markboth{IEEE Transactions on Cloud Computing,~Vol.~, No.~, Mar~2020}%
\markboth{}%
{Lin \MakeLowercase{\textit{et al.}}: Efficient Network Function Backup by Update Piggybacking}

\IEEEtitleabstractindextext{%
\begin{abstract}
  Network Function Virtualization (NFV) and Service Function Chaining
	(SFC) have been widely used to enable flexible and agile network
	management. To enhance reliability, some research has proposed to
	deploy backup function instances for prompt recovery when a
	primary instance fails. While most of the recent studies focus on
	speeding up recovery, less attention has been paid to the problem
	of minimizing the state update cost. In this work, we present
	\name\ ({\em Piggyback-based Backup}), an efficient backup
	instance deployment and update protocol.  Our key idea is to reuse
	the existing service chains traversing through servers in a network
	to help piggyback the update information.  By doing this, we
	eliminate the header overhead and reduce the amount of update
	traffic significantly.  To realize such a piggyback-based update
	more efficiently, we investigate the {\em backup instance
	deployment} and {\em chain selection} problems to enhance
	piggybacking opportunities and reduce the forwarding hop counts
	with explicit consideration of the distribution of service chains.
	Our simulation results show that \name\ reduces the average
	overall update overhead by 47.65\% and 39.56\%, respectively, in a
	fat-tree topology as compared to random deployment and shortest
	path based deployment. 
\end{abstract}

\begin{IEEEkeywords}
Piggybacking, State update, Function backup, Instance deployment
\end{IEEEkeywords}
}

\maketitle

\IEEEdisplaynontitleabstractindextext
\IEEEpeerreviewmaketitle

\IEEEraisesectionheading{
	\secc{Introduction}}

\IEEEPARstart{N}{etwork}
Functions (NFs) typically provide services such as load balancers,
firewalls, Network Address Translation (NAT), and deep packet
inspection in a network. Conventionally, those services can only be
implemented in specific hardware, which is however not flexible and
hardly customized for clients. With the emergence of Software-Defined
Networking (SDN), it has become elastic and efficient to support
software-based services in Virtual Machines (VMs), which is known as
{\em Network Function Virtualization} (NFV)~\cite{7243304}.  The
design of NFV enables network operators to install virtualized NFs and
provide clients customized services by {\em Service Function Chaining}
(SFC).  A client or network application can request a set of network
functions executed sequentially according to its dynamic demands.

For a network enabling NFV, system reliability is of importance since
virtual machines are prone to errors and recovering virtual NFs
usually takes a non-negligible latency~\cite{Stratus1}.  Hence,
providing fault tolerance is essential for an NFV-enabled network.
Some other middlebox
frameworks~\cite{9007021,201545,3323276,2523635,2787501} have
investigated how to provide fault tolerance for individual
middleboxes.  Most of fault-tolerant designs enhance reliability by
introducing
redundancy~\cite{Francisco2017,Carpio2017,Jacobson2014,Nobach2017,Stratus,Kanizo2017}.
In particular, to ensure the reliability of a network function, one
can deploy a primary NF instance in a VM and further install a
redundant NF instance as a backup in a different VM.  Or, two NF
instances can be installed in different VMs, both used to serve their
own clients but, at the same time, acting as the backup of each other.
This type of fault tolerance is called {\em active-active} backup as
both the primary and redundant NF instances are actively used.

However, for active-active backup, primary and redundant NF instances
usually operate independently and do not synchronize their states.
That is, the states will not be consistent when services are migrated
to redundant instances due to the failure of their primary NF
instances.  For example, a service may have been authorized by its
primary NF instance to access a private database but is rejected by
its backup instance as the verification information is not known by
the backup.  To resolve this limitation, an alternative, called {\em
active-standby}
backup~\cite{Jacobson2014,Nobach2017,Stratus,Kanizo2017}, has been
proposed to allow a backup to only keep the state information of
primary NF instances but do not serve clients. Those standby backup
instances can be activated immediately with the up-to-date states once
their primary instances fail.

Recent efforts~\cite{Jacobson2014,Nobach2017,Stratus} have
investigated an efficient and light-weight scheme to launch standby
backup instances. However, an open problem still remains unsolved:
{\em efficient state update}.  To ensure state consistency, each
stateful primary instance should periodically forward state
information to its corresponding backup instance.  Those periodical
state updates would incur tremendous traffic loads. We notice that the
update cost closely depends on the locations of primary and backup
instances. Intuitively, we can reduce the update cost by installing a
backup instance close to its primary instance. However, as the
computing resources of a server are limited, a backup instance may
have to serve multiple primary instances who compete with each other.
Hence, our goal is to investigate an efficient backup instance
deployment and update protocol so as to minimize the overall state
update cost.

In this \article, we present \name\ ({\em Piggyback-based Backup}), an
efficient backup instance deployment and update protocol.  Our design
is motivated by a key observation: {\em a network has typically served
a large number of service function chains that traverse through
diverse routes and can help piggyback the backup update information.}
As state updates usually generate periodical but small packets, the
headers of those small update packets become the major overhead of
state updates.  By piggybacking the state information in existing
SFCs, we can exclude the header for an update and significantly reduce
the update cost, as illustrated in Fig.~\ref{fig:1}.  More
importantly, while a network may admit several service chains over
time, we do not need to select a dedicated chain to piggyback the
update. Instead, any chain with traffic traversing from a primary
instance to a backup instance can be dynamically selected to help
forward the updates.
%If the states of NF A in machine 3 and NF C in machine 4 are backed up
%in machine 5, the solid origin service function chain can help
%piggyback the update information in its payload and carry the update
%to the backup machine, i.e., 5. Consider another example where the
%state of NF E in machine 6 is backed up in machine 3. Since no service
%chain traversing from machine 6 to machine 3, machine 6 should push
%the update information of NF E using stand-along packets, which
%require additional header overhead, to the backup machine, i.e., 3,
%along the dotted green path. 

To realize this idea, we have to address two practical challenges.
First, to enable piggybacking, a {\em piggybackable backup instance}
should be deployed in a VM traversed by existing SFCs who have passed
through its primary instances. For example, for the primary NF B in
node 2 in Fig.~\ref{fig:1}, if its backup function is deployed in node
5, the lower solid orange service chain can help piggyback updates
in its payload and carry the updates from the primary node 2 to the
backup node 5. Though there may exist several chains traversing
through a primary instance and can help piggyback the backup
information, their routing paths, however, have already been assigned
based on their customized demands. Hence, the cost of information
updates does not depend on the distance between the primary instance
and backup instance anymore but is determined by the path length of
piggybacking chains from a primary instance to its backup instance.
Second, since many SFCs can be candidates that help piggyback the
update information, the one with a shorter piggybacking path should be
preferable. However, as traffic arrivals of an SFC are usually
unpredictable, selecting proper chains to piggyback the updates
becomes difficult.  Finally, as VM resources are limited and the
distribution of SFCs cannot be controlled, some primary NF instances
may not have any chance to piggyback their updates, e.g., NF E in node
6 in Fig.~\ref{fig:1}.  Since no service chain traversing from node 6
to an available node, we can only deploy the backup function of NF E
in some available node, e.g., node 3, and push its updates using
stand-along packets, which require additional header overhead, to the
backup node along the dotted green path.

To address those challenges, we develop a backup instance deployment
algorithm that enhances the opportunities of piggybacking and
minimizes the cost of stand-alone updates. Our piggyback-based
placement explicitly considers the locations of primary function
instances and the distribution of chains, as a result allowing backup
instances to be more likely traversed by chains and realizing
piggybacking.  We then propose a chain selection scheme based on chain
arrival prediction to reduce the piggybacking cost. For those
non-piggybackable instances, we further deploy stand-alone backup
instances to ensure full reliability.  

Our contributions are as follows:
\begin{itemize}
	\item We enable piggybacking update by a sophisticated back
		instance deployment algorithm that optimizes piggybacking
		opportunities.
	\item We design a chain selection scheme to identify a short chain
		that is most likely to piggyback the update information on
		time.  The scheme can hence balance the tradeoff between the
		piggybacking cost and the update latency.
	\item We conduct extensive simulations to verify the effectiveness
		of our design in a fat-tree topology.  The simulation results
		show that \name\ outperforms both random deployment and
		shortest path based deployment. The overall update overhead
		can be reduced by about 40\%--50\%. 
\end{itemize}

The rest of the \article\ is organized as follows.
\sect~\ref{sec:related} reviews recent work on NFV fault-tolerant
designs and NF deployment. \sect~\ref{sec:problem} formally defines
the backup instance deployment problem, and \sect~\ref{sec:design}
details the design of \name. We evaluate the performance in
\sect~\ref{sec:results} and conclude this work in
\sect~\ref{sec:conclusion}.

\ifx\isthesis\undefined
        \begin{figure}[t]
            \centering
            \includegraphics[width=3in]{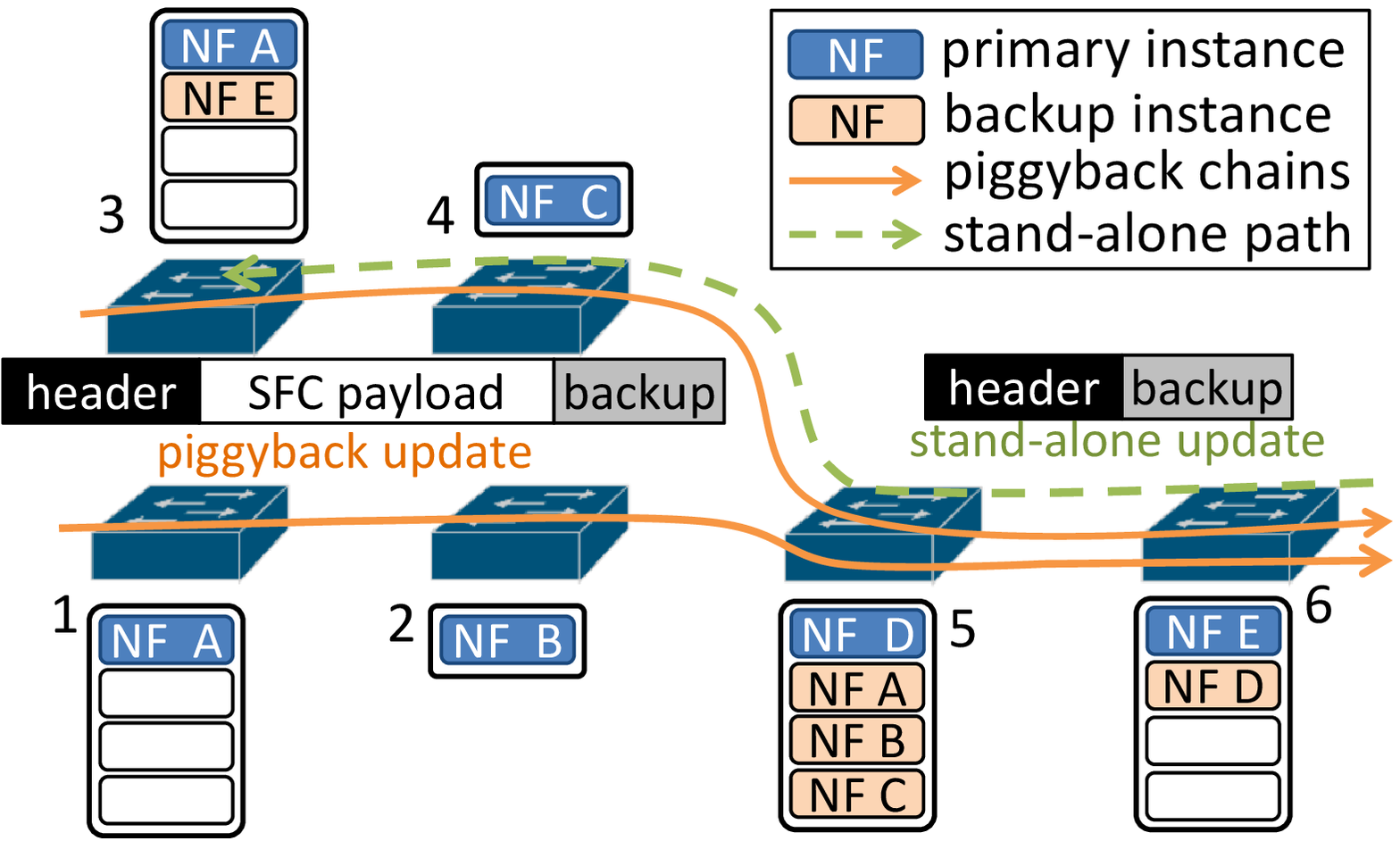}
            \caption{Reuse existing SFCs to piggyback backup updates}
            \label{fig:1}
        \end{figure}
\else
\afterpage{
    \vspace*{\fill}{
        \begin{figure}[ht]
            \centering
            \includegraphics[width=4.5in]{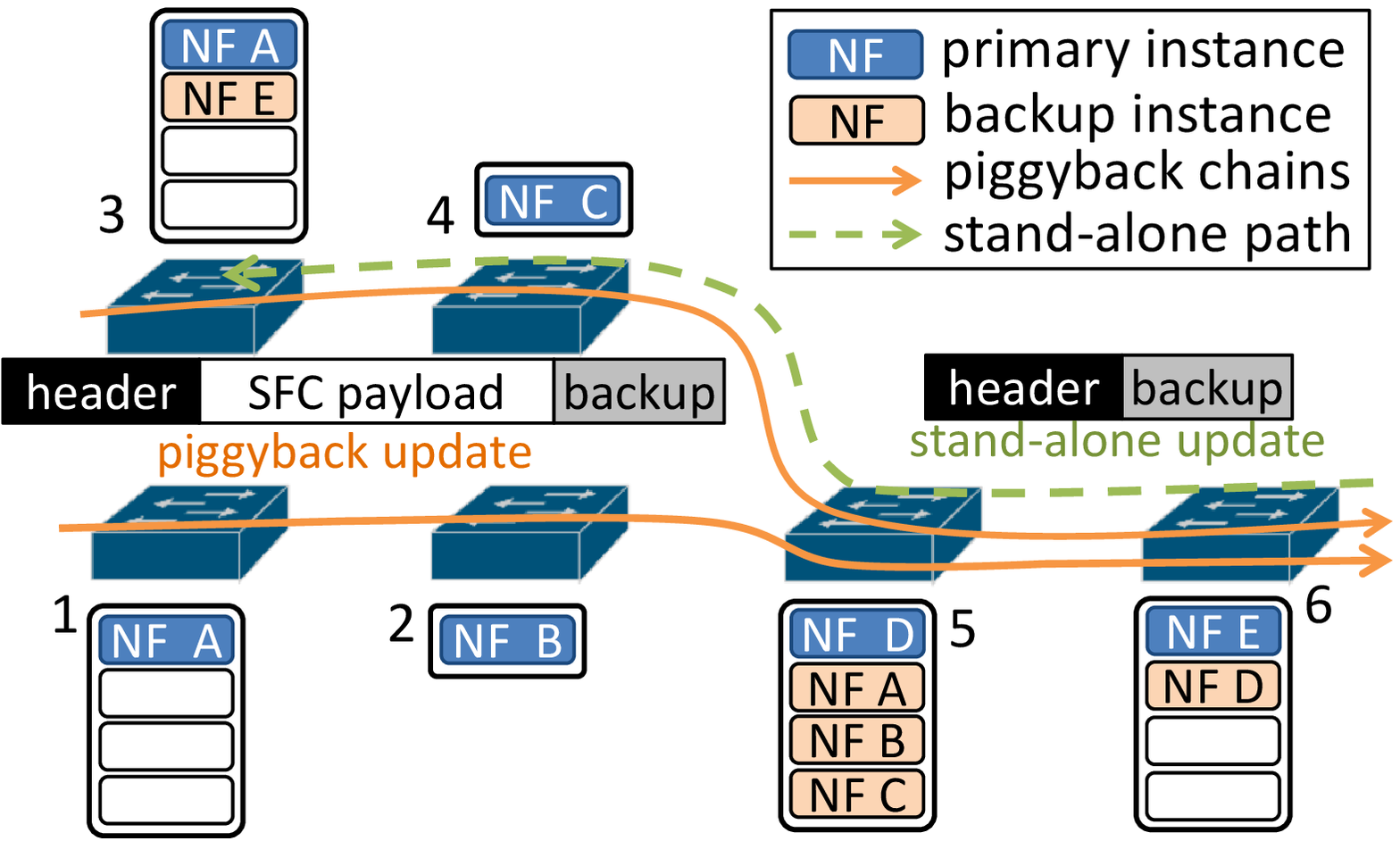}
            \caption{Reuse existing SFCs to piggyback backup updates}
            \label{fig:1}
        \end{figure}
    }
    \vspace*{\fill}
    \clearpage
}
\fi

\secc{Related Work}\label{sec:related}
Recent research related to backup NF deployment falls in three
categories:

\ifx\isthesis\defined
\subsecc{Fault Tolerance Designs}
\else
\vskip 0.05in
\noindent {\bf Fault tolerance design:}
\fi
Several recent studies have investigated how to ensure network
reliability~\cite{6983797,Cully2008,Sherry2015,Rajagopalan2013}.  The
challenges of software reliability assurance for NFV is first
introduced in~\cite{6983797}.  The work \cite{Cully2008} offers a
checkpoint-based fault-tolerant network with three high-level goals:
generality, transparency, and seamless failure recovery.
\cite{Rajagopalan2013} is the first fault-tolerant system for
Virtualized NFs (VNFs). It maintains the states of network flows in
different NF instances by using light checkpoints.  In general,
reliability can be supported by two different ways:
active-active~\cite{Francisco2017,Carpio2017} or
active-standby~\cite{Jacobson2014,Nobach2017,Stratus,Kanizo2017}.  The
active-active solutions~\cite{Francisco2017,Carpio2017} deploy
multiple NF instances to ensure a chain to be migrated to a different
NF instance when its primary instance fails. All NF instances should
support their own chains and be used as a backup at the same time.
They hence require the same level of computing resources. 
%\del{All NF instances should support its own chains and be used as a backup at the same time. They hence require the same level of computing resources.}
However, there exists no state transferring among different instances,
i.e., no state consistency guaranteed. The active-standby solutions
deploy lightweight backup instances, which only keep the states of
primary functions but do not serve SFCs. It hence requires negligible
computing resources and guarantees state consistency. The
work~\cite{Kanizo2017} adopts this solution and deploys backup
instances to optimize the reliability of the whole network without
considering the update cost.  Our work also adopts active-standby
backup but aims at minimizing the update overhead.

\ifx\isthesis\defined
\subsecc{State Maintenance}
\else
\vskip 0.05in
\noindent {\bf State maintenance:}
\fi
State consistency can be maintained by three types of
implementation~\cite{Stratus}: hardware~\cite{Potharaju2013},
VNFs~\cite{Chandrasekaran2016}, and
software~\cite{Jacobson2014}\cite{Nobach2017}\cite{Stratus}.
Hardware-based implementation~\cite{Potharaju2013} is the most
traditional method, which is more efficient but cannot be customized.
The VNF-based implementation allows network managers to flexibly
program network functions that customize how the states should be
protected and maintained. For instance, the
work~\cite{Chandrasekaran2016} leverages a state transformation
technique to avoid the need for manual intervention. Modifying NF
programs, however, requires a high manpower cost and makes version
control difficult.  Finally, the software-based implementation is the
most commonly used solution nowadays. The
systems~\cite{Jacobson2014}\cite{Nobach2017} implement a robust state
transferring scheme for SDNs without modifying the NFs.
Stratus~\cite{Stratus} is a cloud-based solution developed to achieve
state consistency in stateful VNFs. Due to the popularity of
software-based solutions, we adopt this type of implementation and
investigate how to deploy backup NFs so as to reduce the state update
cost subject to the reliability requirement.

\ifx\isthesis\defined
\subsecc{Primary NF Instance Deployment}
\else
\vskip 0.05in
\noindent {\bf Primary NF instance deployment:}
\fi
The latency of service function chaining is closely related to how NF
instances are deployed.  Recently, a large number of studies have
devoted to enhancing efficiency and resource utilization of NF
deployment in an
SDN~\cite{Alicherry2012,Jayasinghe2011,Meng2010,Addis2015,Zhang2013,Tomassilli2018,Mehraghdam2014,Li2018,Luizelli2015,Hirwe2016}.
The work \cite{Luizelli2015} derives an integer linear programming
model to solve the joint NF placement, assignment, and routing
problems. It first identifies suitable positions to deploy VNFs and
then assigns proper instances to the NF requests of a chain subject to
a given latency constraint.  LightChain~\cite{Hirwe2016} determines
VNF placement locations to minimize the number of hop counts of every
SFC but does not consider resource utilization. While the objective of
primary NF deployment is to improve the efficiency of service
chaining, the objective of backup NF deployment is however to ensure
high reliability and low update costs, which are the focus of our
work.

 %\newpage
\secc{Problem Statement and System Model}\label{sec:problem}
\begin{comment}
We first define the backup instance deployment problem and give a
motivating example demonstrating the intuition of our piggyback-based
design.
\end{comment}
In this \sectlower, we first formally define the backup instance
deployment problem and then express the system model of \name.

%\subsecc{Problem Statement}
\subsecc{Problem Statement}
An SDN network is represented as a directed graph
$G=(\mathcal{V},\mathcal{E})$, where $\mathcal{V}$ is the set of
physical machines (servers) and $\mathcal{E}$ is the set of network
edges.  The network serves a set of function types $\mathcal{F}$ and
installs multiple instances for every type $f\in \mathcal{F}$ in
different physical servers. We collect the set of deployed primary
function instances as a set $\mathcal{N}$, and, for simplicity,
$\mathcal{N}_f\subseteq \mathcal{N}$ represents the set of instances
of function type $f$, i.e., $\cup_f \mathcal{N}_f = \mathcal{N}$.  Let
$\mathcal{C}$ be the set of currently served service chains.  We
assume that the deployment of primary function instances is given and
the instances serving each SFC $c\in \mathcal{C}$ have also been
assigned. Let notations $u(n)$ and $f(n)$, respectively, denote the
server $v\in V$ that installs the primary function instance $n \in
\mathcal{N}$ and the function type of $n$. Our goal is to install a
backup function instance for every primary function instance $n \in
\mathcal{N}$ in a physical server $v \in \mathcal{V}$ different from
$u(n)$.

Since each physical server has limited capacity, we assume that each
$v\in \mathcal{V}$ can at most install $B_v$ backup function
instances. We define the set of backup instances installed in $v$ as
$\mathcal{B}_v$, and, hence, the set of all backup instances is
$\mathcal{B}=\cup_{v\in \mathcal{V}} \mathcal{B}_v$.  Due to limited
resources, we assume that each backup function instance $b\in
\mathcal{B}$ can backup the states for multiple primary function
instances of the same type. However, each backup instance can support
at most $K$ primary instances. Each primary instance $n\in
\mathcal{N}$ will be assigned a backup instance and periodically send
the up-to-date states to its associated backup instance. The update
cost of an instance $n$, denoted by $w_n$, is defined as the
aggregated number of bits over all network edges per update. That is,
if the update of an instance $n$ including $L$ bits traversing through
$k$ hops in the network, the total update cost $w_n$ will equal to
$w_n=L*k$.  We aim at identifying the optimal backup instance
deployment such that the overall update forwarding cost can be
minimized.

\subsecc{System Model}

In our system, each function type $f$ can be backed up in many servers,
but each primary function instance $n$ of the function type $f$ should
be associated exactly with one backup server. A backup server of
function type $f$ can be associated with up to $K$ primary instances.
Our goal is to jointly solve the problems of backup instance
deployment and primary-backup instance association.  The two problems
can be represented by the following binary decision variables: 
\[
I_{f,v}=
\left\{
\begin{array}{ll}
	1,&\text{ if function type } f\in\mathcal{F} \text{ backed up in }
	v\in\mathcal{V},\\
	0,&\text{ otherwise}, \text{ and}
\end{array}
\right . 
\]
\[
J_{n,v}=
\left\{
\begin{array}{ll}
	1,&\text{if primary instance } n\in\mathcal{N} \text{ backed up in
	} v\in\mathcal{V},\\
	0,&\text{otherwise.}
\end{array}
\right .
\]

By identifying a proper configuration of the above two variables, we
can create more piggyback opportunities, as a result reducing the
update cost. The notations are summarized in Table \ref{table:1}.
Such a piggyback-based backup instance deployment and assignment
problem can be formulated as the following integer linear programming
(ILP) problem:
\begin{subequations}\label{eq:1}
    \begin{align}
        \label{eq:1a}
		\min \sum_{n\in\mathcal{N}} w_n = \min \sum_{n \in \mathcal{N}}
		w_n^{\text{pg}} + w_n^{\text{alone}}
    \end{align}
    \textrm{subject to:}
    \begin{align}
        \label{eq:1b}
		\sum_{f\in\mathcal{F}} I_{f,v} \le B_v, \forall v\in
		\mathcal{V}
    \end{align}
    \begin{align}
        \label{eq:1c}
		\sum_{v\in \mathcal{V}} J_{n,v} = 1, \forall n\in\mathcal{N}
    \end{align}
    \begin{align}
        \label{eq:1d}
		\sum_{n\in\mathcal{N}_f} J_{n,v} \le K*I_{f,v}, \forall
		v\in\mathcal{V}, f\in\mathcal{F} 
    \end{align}
    \begin{align}
        \begin{split}
        \label{eq:1f}
			w_n^{\text{pg}}=\sum_{v\in \mathcal{V}} J_{n,v} \left(\sum_{c \in
			\mathcal{C}_{u(n),v}}l_{c,u(n),v}\omega^{\text{pg}}/|\mathcal{C}|\right),
			\forall n\in\mathcal{N}
        \end{split}
    \end{align}
	\begin{align}
        \begin{split}
        \label{eq:1g}
			w_n^{\text{alone}}{=}\sum_{v\in \mathcal{V}} J_{n,v}\left(|\mathcal{C}
			\backslash \mathcal{C}_{u(n),v}|l^{\min}_{u(n),v}
			\omega^{\text{alone}}/|\mathcal{C}|\right),{\forall}n{\in}\mathcal{N}
        \end{split}
    \end{align}
	\begin{align}
		I_{f,v}\in \{0,1\}, J_{n,v}\in\{0,1\}
	\end{align}
\end{subequations}

\begin{table}[t] 
		\caption{List of Notations}
		%\vspace{5mm}
		\centering  \label{tab:ParameterSetting}
		\begin{tabular}{lp{6cm}}
			\toprule
			       Notation        & Definition  \\ 
			\toprule
				   $G=(\mathcal{V},\mathcal{E})$      & network graph
				   including a set of physical servers \\
			       $\mathcal{F}$      & set of function types \\
			       $\mathcal{N}$      & set of installed primary function instances \\
				   $u(n)$ & server that installs instance $n$ \\
				   $f(n)$ & function type of instance $n$ \\
			       $\mathcal{C}$      & set of existing chains \\
				   $\mathcal{C}_{u,v}$      & set of existing chains
				   traversing from $u$ to $v$\\
			       $l_{c,u,v}$     & number of hops traversed by chain $c$ from $u$ to $v$ \\
			       $l^{\min}_{u,v}$     & shortest path length from $u$ to $v$ \\
				   $\omega^{\text{pg}}$     & cost of a piggyback update
				   message \\
				   $\omega^{\text{alone}}$    & cost of a stand-alone
				   update message\\
				   $\mathcal{B}_v$ & set of backup instances installed
				   in server $v$\\
			       $B_v$    & maximal number of backup instances
				   allowed in $v$\\
			       $K$    & maximal number of primary instances a
				   backup instance can support\\
			 \bottomrule
		\end{tabular}
		\label{table:1}
\end{table}

We assume that each backup server has a limited computation power and
storage. Hence, Eq.~\eqref{eq:1b} limits the number of backup
instances deployed in $v$ by its capacity $B_v$. Eq.~\eqref{eq:1c}
requires each primary function instance $n$ to be assigned exactly one
backup node. Then, Eq.~\eqref{eq:1d} constrains a primary instance $n$
of type $f$ to be associated with a backup node $v$ only if $v$ has
installed a backup instance of its function type, i.e., $I_{f,v} = 1$.
Otherwise, the primary instance $n$ cannot associate with $v$ when
$I_{f,v}=0$.  Also, this expression forces the number of primary
instances associated with a server $v$ to be no more than its capacity
$K$. In our design, we allow a service chain traversing through $u(n)$
to $v$ to help piggyback the update information from primary instance
$n$ to its backup node $v$. Let $\mathcal{C}_{u(n),v}$ represent the
set of those piggybacking chains.  Since different chains traverse
through different paths, the piggybacking cost of chain $c\in
\mathcal{C}_{u(n),v}$ can be calculated by
$l_{c,u(n),v}\omega^{\text{pg}}$, as expressed in Eq.~\eqref{eq:1f},
where $l_{c,u(n),v}$ is the number of hops traversed by chain $c$ from
$u(n)$ to $v$. If, unfortunately, no chain can help piggyback the
update information, we should send stand-alone update packets, each of
which costs $l^{\min}_{u(n),v}\omega^{\text{alone}}$ as in
Eq.~\eqref{eq:1g}, where $l^{\min}_{u(n),v}$ is the shortest path
length between $u(n)$ and $v$. We further normalize the two costs
$w_n^{\text{pg}}$ and $w_n^{\text{alone}}$ in
Eqs.~(\ref{eq:1f},\ref{eq:1g}) by the cardinality of $\mathcal{C}$ to
derive the expected cost. Finally, the objective in Eq.~\eqref{eq:1a}
aims at minimizing the overall update cost.

The optimal solution of the above ILP problem can be solved by
exhaustive search or some generic algorithm, e.g., branch and
bound~\cite{Zahra2015}, which however introduces a high complexity.
We hence focus on developing a more practical light-weight greedy
algorithm that identifies an efficient and near-optimal backup
strategy with a joint consideration of the update cost and resource
utilization.  Our greedy algorithm contains two modes, {\em piggyback
mode} and {\em stand-alone mode}. The piggyback mode deploys as many
backup instances that can be traversed by  SFCs as possible. It
further identifies a suitable chain to piggyback the updates with small
costs. For the remaining primary instances who cannot be supported by
piggyback-based backup, the stand-alone mode then installs backup
instances in the remaining available servers for reducing their update
costs.

\secc{PiggyBackup Design}\label{sec:design}
We now describe the proposed \name\ design that deploys piggyback
backup instances to enhance piggybacking opportunities
(\sect~\ref{sec:piggyback}) and selects proper piggybacking service
chains (\sect~\ref{sec:chain}).  We then explain how to deploy the
remaining stand-alone backup instances to ensure full coverage
(\sect~\ref{sec:shortest}).

\subsecc{Piggyback Instance Deployment}\label{sec:piggyback}
Since the opportunities of piggybacking updates are closely correlated
to how service chains traverse, we hence propose a greedy algorithm to
deploy backup instances based on the distribution and loading of
service chains, as summarized in Algorithm \ref{algo:1}.  Since each
server has a limited capacity, we should deploy a backup instance of
type $f$ on a server $v$ if more service chains can traverse from
different primary instances of the same type to $v$. To this end, we
collect all the chains $c\in\mathcal{C}$ that traverse through any
instance $n\in\mathcal{N}_f$ of type $f$ as a set $\mathcal{C}_f$ and
deploy backup function instances for various function types $f$ in
descending order of $|\mathcal{C}_f|$ (line 2).

\begin{algorithm}[t]
	\SetKwInOut{Input}{Input}
	\SetKwInOut{Output}{Output}
	\Input{ $G$: network graph;  $\mathcal{F}$: set of NF types;
	$\mathcal{N}_f$: set of primary instances of function type
	$f\in\mathcal{F}$; $\mathcal{C}$: set of SFCs; $K$: backup limit;
	$B_v$: capacity of server $v$}
	%\State{D=Floyd\_Warshall($T$)}
	\While{$\mathcal{F}\neq\emptyset$}{
		$C_f\leftarrow$ set of chain including function type $f$\\
		$f\leftarrow \arg\max_f |\mathcal{C}_f|$\\
		$\mathcal{F} \leftarrow \mathcal{F}\backslash\{f\}$\\
		\While{$\mathcal{N}_f\neq\emptyset$}{
			$\lambda_{v,n} \leftarrow \sum_{c\in \mathcal{C}_f}r_c/l_{c,u(n),v},
			\forall n\in\mathcal{N}_f, v\in\mathcal{V}, B_v > 0$\\
			$\lambda_v \leftarrow
			\sum_{n\in\mathcal{N}_f}\lambda_{v,n}, \forall
			v\in\mathcal{V}, B_v > 0$\\
			{\bf if } $\max_{v\in\mathcal{V}, B_v>0} \lambda_v = 0$
			{\bf then break;}\\
			\textsf{\footnotesize{// select a node with highest
			score to back up for $f$}}\\
			$v^* \leftarrow\arg\max_{v} \lambda_v$\\
			install backup instance in $v^*$, $B_{v^*} \leftarrow B_{v^*}-1$\\
			$I_{f,v^*} \leftarrow 1$\\
			\textsf{\footnotesize{// associate top $K$ instances
			with backup instance in $v^*$}}\\
			sort $n\in\mathcal{N}_f$ in descending order of
			$\lambda_{v*,n}$\\
			$J_{n_k,v^*} \leftarrow 1, k = 1, 2, \cdots,K$\\
			remove $n_1,\cdots,n_K$ from $\mathcal{N}_f$\\
			%remove chains that covered backed up instances from $C_f$\\
			%$C_f=C_f-C_{covered}$\\
			%$P=P-P_{covered}$\\
			}
			}
		{\bf return} $I_{f,v}, J_{n,v}, \forall  n \in
		\mathcal{N}_f, f\in\mathcal{F},v\in\mathcal{V}$
			\caption{Piggyback backup instance deployment}\label{algo:1}
\end{algorithm}

To identify a proper server $v$ to install the backup instance of type
$f$, we try to quantify the amount of potential traffic that can
piggyback updates of any primary instance $n\in\mathcal{N}_f$ to
server $v$.  For simplicity, we call the traffic of any $c\in
\mathcal{C}_f$ that arrives a backup server $v$ after being served by
an instance $n\in\mathcal{N}_f$ {\em piggybackable traffic} to server
$v$ for short.  To create more piggyback opportunities, we should
deploy a backup instance in a server $v$ traversed by more {\em
piggybackable traffic}.  However, the cost of piggybacking updates
also depends on the distance between the location of a primary
instance and server $v$.  Hence, to quantitatively prioritize all the
servers available for installing a backup instance, we sort the
available servers in descending order of the following performance
metric (lines 5-6):
\begin{eqnarray}
	\lambda_v &=& \sum_{n\in\mathcal{N}_f}\lambda_{v,n}, \text{where}\\
	\lambda_{v,n}&=&\sum_{c\in\mathcal{C}_f}\frac{r_c}{l_{c,u(n),v}},
\end{eqnarray}
and $r_c$ is the average traffic rate of chain $c$ and $l_{c,u(n),v}$
is the length of the segment traversed by chain $c$ from the server of
its primary instance $u(n)$ to a potential backup server $v$. Note
that $l_{c,u(n),v}$ is set to $\infty$ if chain $c$ does not arrive
$v$ after $u(n)$, i.e., not piggybackable.  At a high level, the
metric $\lambda_{v,n}$ represents the sum traffic arrival $r_c$ of
chain $c$ passing throughput a server $v$ divided by the path length
of piggybacking $l_{c,u(n),v}$, while the metric $\lambda_v$ is total
value summed over all possible instances $n \in \mathcal{N}_f$.  That
is, a server with more piggybackable traffic and a shorter piggyback
path length will result in a larger score $\lambda_v$.  Hence, we
deploy the backup instance for type $f$ in the server $v^*$ with the
maximal score $\lambda_{v^*}$ (lines 9-11), i.e., $I_{f,v^*}=1$. If no
such server exists, we proceed to the next function type $f$ (line 7).

Next, recall that a backup instance can support at most $K$ primary
instances. We should further associate the backup instance in $v^*$
with $K$ primary instances.  To this end, we exploit the same design
intuition to identify the primary instances that produce most
piggybackable traffic.  In particular, we sort primary function
instances $n\in\mathcal{N}_f$ in descending order of their
contributions $\lambda_{v^*,n}$ (line 13) and select the instances
with the top-$K$ scores $\lambda_{v^*,n}$, denoted by $n_1,
n_2,\cdots, n_K$, to associate with the backup instance installed in
server $v^*$, i.e., $J_{n_k,v^*}=1, k=1,2,\cdots,K$ (lines 14--15).
The above deployment and association procedure repeats until every
primary instance has been associated with a piggyback backup instance
or all the servers available for piggybacking have been fully
occupied.

\begin{figure}[tb]
    \centering
    \includegraphics[width=3.3in]{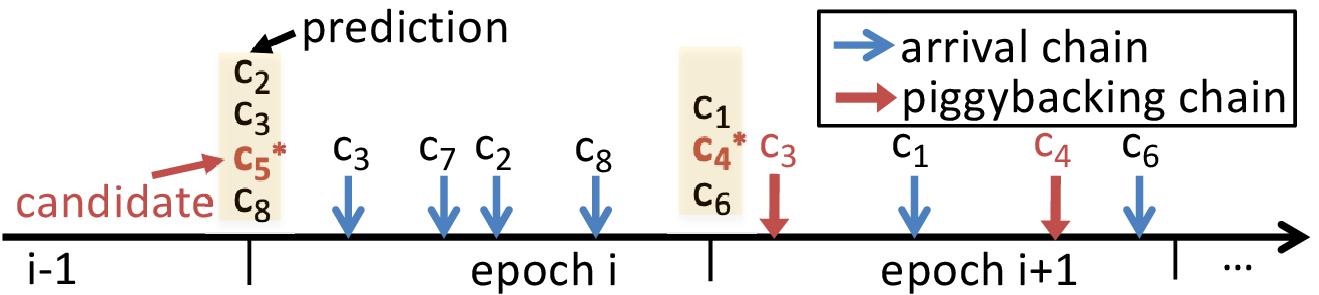}
    \caption{Example of piggybacking chain selection}
    \label{fig:pg_chain}
\end{figure}

\subsecc{Piggybacking Chain Selection}\label{sec:chain}
Once piggybackable backup instances have been deployed, there may be
several chains that can help piggyback the update information since
each packet of any service chain traversing from a primary instance to
a backup instance can help piggyback the update.  That is, there is no
need to select a dedicated chain to piggyback the updates. Instead,
any chain with packets traversing through the backup instance can help
forward the updates. This design is especially practical for a real
network where service chains may join and leave dynamically.  An
intuitive strategy is piggybacking the periodical update information
in a packet of any chain, i.e., first-come-first-serve (FCFS).
However, different chains usually go through different function
instances and, thereby, traverse through various path lengths. To save
the traffic load for the entire network, a more reasonable strategy is
to ask the shortest chain to help piggyback periodical updates.  This
alternative, however, faces another issue: the traffic arrival rate of
the shortest chain may not be high enough to deliver periodical
updates on-time. As different chains typically generate heterogeneous
traffic rates along paths of different lengths, we aim at selecting
piggybacking chains with consideration of both the update interval and
the update cost.

As packet arrivals of a chain are usually unpredictable, we propose a
{\em bounded-waiting} chain selection scheme. In particular, we
partition time into epochs, each of which has a fixed interval $T$
equal to the update interval.  At a high level, in the beginning of
epoch $i$, each primary function instance predicts whether a chain
would arrive in epoch $i$ and identifies the chain with the shortest
path to its backup instance as the candidate $c^*$ from those who are
likely to arrive in epoch $i$.  The primary instance then waits for
the packet of the candidate $c^*$ for piggybacking the update. If
$c^*$ indeed arrives in epoch $i$, the information can be smoothly
piggybacked to the backup instance. If not, the primary instance will
embed the update in the first packet of any chain in epoch $i+1$,
i.e., FCFS. If, unfortunately, there is no packet arrival in an epoch,
a primary function can send a stand-alone update to its backup
instance.  By doing this, we bound the waiting time within the epoch
interval $T$ and hence balance the tradeoff between the update cost
and the update latency.

To estimate the arrival pattern of a chain, we assume that packet
arrivals of a chain follow a Poisson process, i.e., the inter-packet
time following the exponential distribution with mean $1/\lambda$,
where $\lambda$ is the mean traffic arrival rate and can be measured
from historical packets. Hence, the primary instance can predict the
arrival time of a chain based on the arrival time of its previous
packet and the mean inter-packet time. Note that even if the packet
arrival pattern does not follow the Poisson process, we can still
measure the empirical mean packet arrival rate via packet sampling and
INT technologies.  With the prediction of traffic arrival patterns,
the primary instance can collect the chains whose next packet would
arrive in epoch $i$ and find its candidate $c^*$ in the beginning of
epoch $i$.  Fig.~\ref{fig:pg_chain} illustrates an example of our
bounded-waiting chain selection scheme. In epoch $i$, the primary
function predicts that chains $c_2, c_3, c_5$, and $c_8$ may have
packet arrivals and selects $c_5$ (with the shortest piggyback length)
as its candidate. It then waits for $c_5$'s arrival in epoch $i$.
However, chain $c_5$, in fact, does not arrive in epoch $i$ as
expected.  Then, the primary instance embeds the update information in
the first packet of epoch $i+1$ (i.e., chain $c_3$). It again performs
the same prediction and selects $c_4$ as the candidate in the
beginning of epoch $i+1$.  This time, $c_4$ arrives in epoch $i+1$ as
expected.  Therefore, the primary instance can ask $c_4$ to piggyback
the next update information.

\begin{figure}[tb]
    \centering
    \includegraphics[scale=0.45]{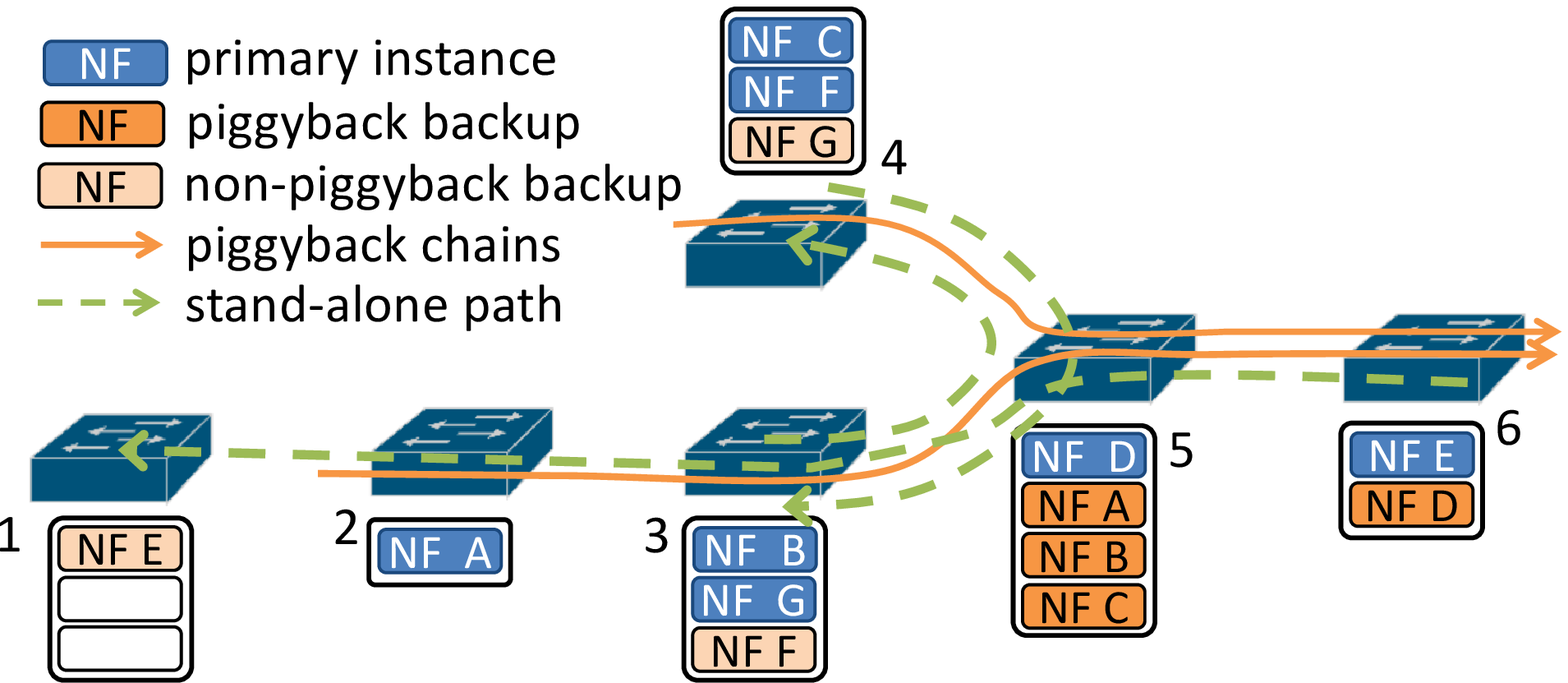}
    \caption{Example of stand-alone backup deployment}
    \label{fig:compete}
\end{figure}

\subsecc{Stand-Alone Instance Deployment}\label{sec:shortest}
After piggyback-mode deployment, some primary instances, denoted by
$\mathcal{N}_{\text{alone}}$, may not be able to find any chain
traversing from it to any server available for backup.  To cover those
remaining primary instances, we will deploy their backup instances in
the available servers and send the updates using stand-alone packets.
Note that an existing backup instance deployed in the piggyback mode
may still be available if it serves fewer than $K$ primary instances.
That is, those backup instances can be reused by stand-alone primary
instances. For a function type $f$, the set of its available servers
can be defined as 
\begin{equation}
	\mathcal{V}_f = \{v:
\sum_{f'\in\mathcal{F}}I_{f',v} < B_v || (I_{f,v}\text{\&}\sum_{n\in\mathcal{N}_f}
J_{n,v} < K)\}.
\end{equation}

An intuitive deployment is to place a backup instance as close to its
serving primary instance as possible. However, the problem is not that
simple since {\em i)} a backup instance is usually shared by several
primary instances, and {\em 2)} multiple primary instances compete for
limited server resources.  Consider Fig.~\ref{fig:compete} as an
example. Assume that servers 1, 3, 4, and 6 can only install one
backup instance. If primary instances NFs G and F in server 3 and
server 4, respectively, both configure their backup instances to the
closest available servers, i.e., server 4 and server 3, they will run
out of the resources nearby the last primary instance, NF E in server
6. Then, the only available remaining server, server 1, will be very
far from the last primary instance NF E, leading to a large update
cost for NF E.  However, if NF F in server 3 can backup in server 1,
NF E in server 6 should be able to backup in server 4 to reduce the
overall update cost.

\begin{algorithm}[t]
\SetKwInOut{Input}{Input}
\SetKwInOut{Output}{Output}
\Input{$\mathcal{N}_{\text{alone}}$: set of  stand-alone primary
	instances; $I_{f,v}$: piggyback-mode deployment}
	$\mathcal{V}_f \leftarrow \{v: \sum_{f'\in\mathcal{F}}I_{f',v} < B_v ||
	  (I_{f,v}\text{\&}\sum_{n\in\mathcal{N}_f} J_{n,v} < K)\},
	  \forall f\in\mathcal{F}$\\
	  \While{$\mathcal{N}_{\text{alone}}\neq \emptyset$} {
		$\Delta_n \leftarrow
		l^{\text{2nd}{\min}}_n-l^{\min}_n,\forall n\in
		\mathcal{N}_{\text{alone}}$\\
		\textsf{\footnotesize{// identify primary instance with
		highest priority}}\\
		$n^*\leftarrow \arg\max_n \Delta_n$\\
		$\mathcal{N}_{\text{alone}} \leftarrow
		\mathcal{N}_{\text{alone}}\backslash\{n^*\}$\\
		{\bf if} $\mathcal{V}_{f(n^*)} = \emptyset$ {\bf then
		continue} \textsf{\footnotesize // no available server}\\
		{\bf if} $I_{f(n^*),v^*}=0$ {\bf then} install backup instance in
		$v^*$\\
		\textsf{\footnotesize{// associate $n^*$ with closest server
		$v^*$}}\\
		$I_{f(n^*),v^*}\leftarrow 1, J_{n^*,v^*}\leftarrow 1$\\
		update the available server set $\mathcal{V}_{f(n^*)}$ \\
		update the costs $l^{\min}_n$ and
		$l^{\text{2nd}{\min}}_n,\forall n\in\mathcal{N}$
		\\
	}
  \caption{Stand-alone backup instance deployment}
  \label{algo:2}
\end{algorithm}

To resolve this issue, the remaining primary instances should
cooperate with each other and identify their backup instance
positions to minimize the overall update cost.  Algorithm~\ref{algo:2}
summarizes our cooperative stand-alone instance deployment scheme.
Intuitively, a primary instance close to more available servers has
more options and, hence, should make its decision later. On the
contrary, a primary instance only close to a few available servers
should get a higher priority to install its backup instance.  

To realize this idea, we quantify the priority of each remaining
primary instance $n \in \mathcal{N}_{\text{alone}}$ by the {\em
additional update cost} $\Delta_n$, which is defined as the difference
between its shortest path length and its second shortest path length
to any available server in $\mathcal{V}_{f(n)}$ (line 3).  For
example, if the lengths of the shortest path and the second shortest
path are $k$ and $k'$, respectively, the additional update cost
$\Delta_n$ will be $(k' - k) * K$ for an update packet of $L$ bits.
Specifically, a primary instance with a smaller additional update cost
will be assigned a lower priority since its update cost can still be
low even if its backup instance locates in the second closest server.
Hence, we sort the primary instances in $\mathcal{N}_{\text{alone}}$
in descending order of their $\Delta_n$ (lines 4-6) and sequentially
associate them with their closest available server $v^*$ in order
(lines 9-10).  If there has existed a non-overloaded backup instance
in $v^*$, it can directly be reused by the associated primary
instance. Otherwise, we deploy a new backup instance in $v^*$ to serve
the associated primary instance (line 8).  The server $v^*$ will be
removed from the available set $\mathcal{V}_f$ if it has been fully
occupied after an assignment. The above procedure repeats until all
the primary instances have been associated with a backup instance or
none of the servers is available (lines 2, 7).

%Take Fig. \ref{fig:3} to explain the problem of naive shortest method,
%let\textceltpal s assume we want to look for backup positions for SF C
%in server 1 and SF A in server 2. If we first select SF C in server 1
%to find a backup position, it will choose server 3 to back up, and
%because server 3 has run out of capacity, SF A in server 2 will choose
%server 8 to back up, the update path will be red in the figure, and
%the hop counts will be 4. Conversely, SF A in server 2 will choose
%server 3 to backup, and SF C in server 1 will choose server 7 to
%backup, the update path will be blue in the figure, and the hop counts
%will be 3.  This example shows that which primary instance back up
%first will affect the final result.

\ifx\isthesis\undefined
        \begin{table}[t] 
		\small
		\caption{Default simulation Setups}
		%\vspace{5mm}
		\centering  \label{tab:Parameter}
		\begin{tabular}{ll}
			\toprule
			       Parameter         & Values  \\ 
			\toprule
					Number of servers &20\\
			       Number of function types      & 20 \\
			       Number of chains      & 50 \\
			       Primary capacity per server & 8 \\
			       Backup capacity per server & 3 \\
				   Backup-primary association limit ($K$) & 5\\
			 \bottomrule
		\end{tabular}
		\label{tb:setting}
\end{table}
\else
%\afterpage{
%    \vspace*{\fill}{
        \begin{table}[t] 
		\small
		\caption{List of Settings}
		%\vspace{5mm}
		\centering  \label{tab:Parameter}
		\begin{tabular}{ll}
			\toprule
			       Parameter         & Default configuration  \\ 
			\toprule
					Number of servers &20\\
			       Number of function types      & 20 \\
			       Number of chains      & 50 \\
			       Primary capacity per server& 8 \\
			       Backup capacity per server & 3 \\
				   Backup-primary association limit ($K$) & 5\\
			 \bottomrule
		\end{tabular}
		\label{tb:setting}
\end{table}
%    }
%    \vspace*{\fill}
%    \clearpage
%}
\fi

\ifx\isthesis\undefined
\begin{figure}[t]
	\centering
	\begin{tabular}{c}
		%\hspace{-12pt}
		\epsfig{width=2.8in,file=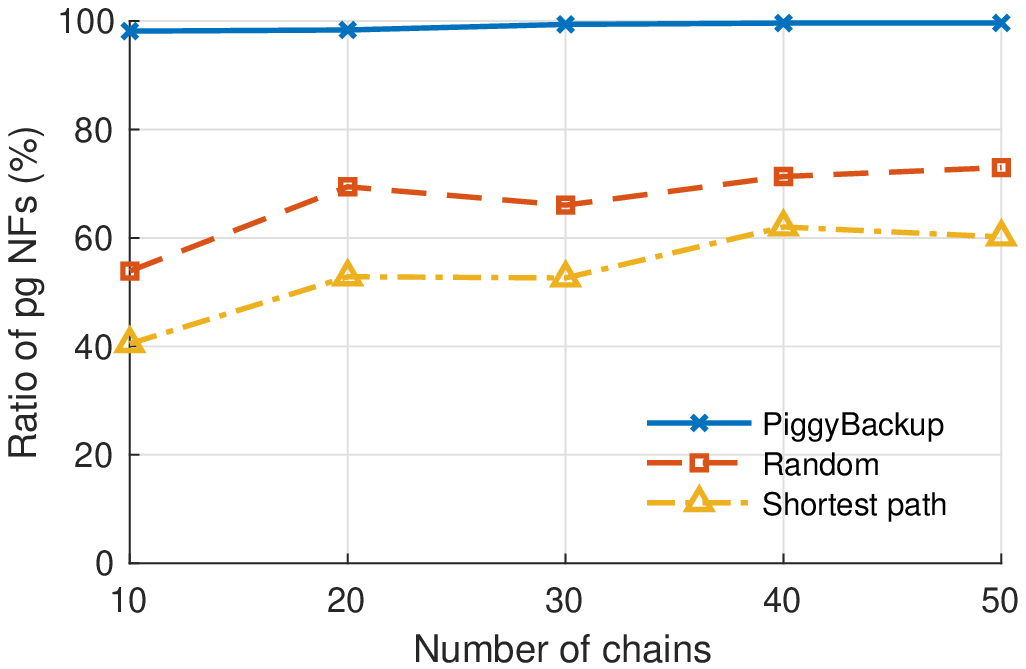}\\
		\vspace{6pt}
		\footnotesize{(a) percentage of piggybackable
		instances} \\
		%\hspace{-10pt}
		\epsfig{width=2.8in,file=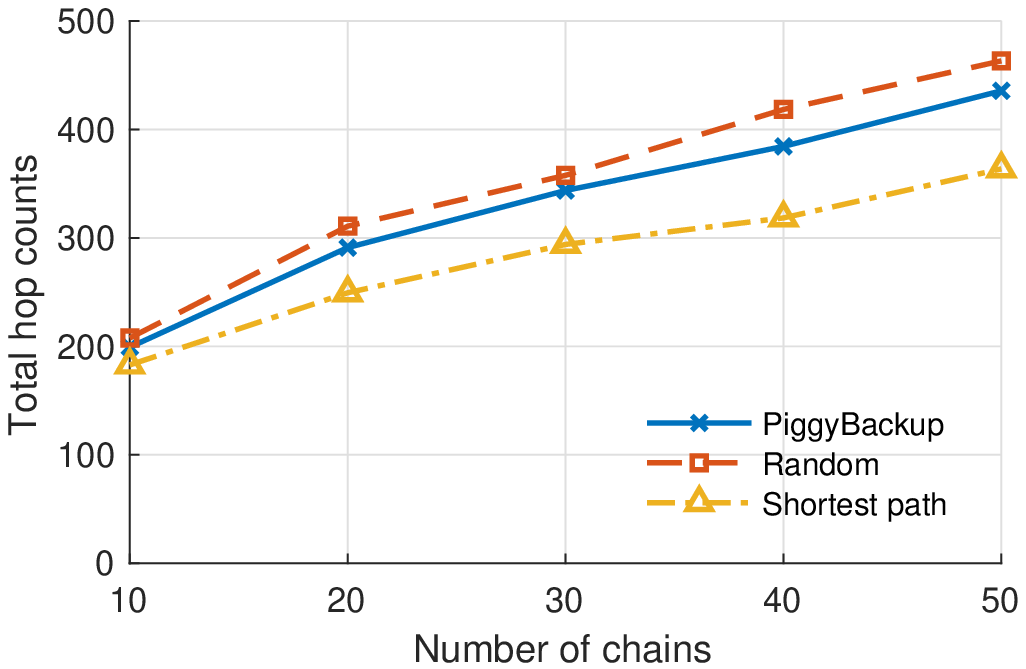}\\
		\vspace{6pt}
		\footnotesize{(b) total number of hop counts} \\
		%\hspace{-10pt}
		\epsfig{width=2.8in,file=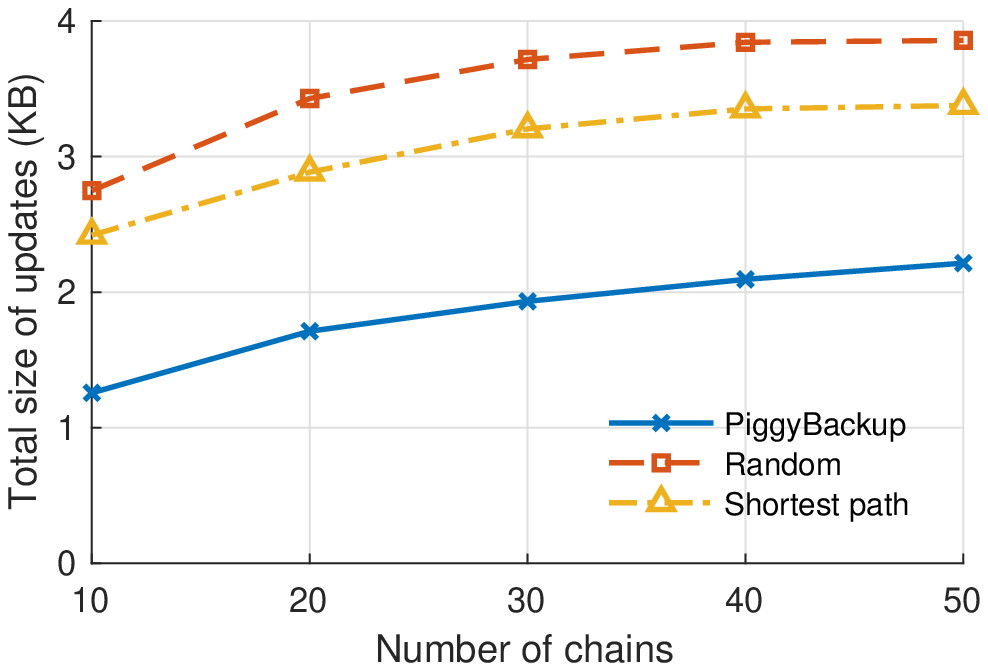}\\
		\footnotesize{(c) total size of updates}
	\end{tabular}
	\caption{Impact of number of chains.}
	\label{fig:chainNum}
\end{figure}
\else
\afterpage{
	\vspace*{\fill}{
		\begin{figure}[ht]
			\centering
			\begin{tabular}{c}
				\epsfig{width=2.8in,file=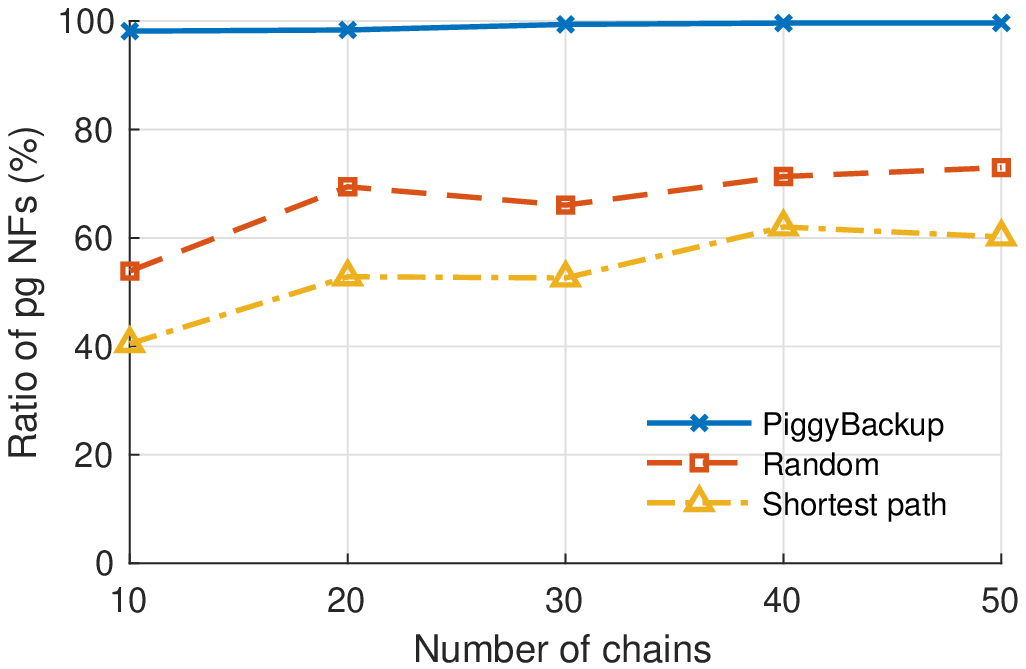}\\
				(a) percentage of piggybackable instances \\
				\epsfig{width=2.8in,file=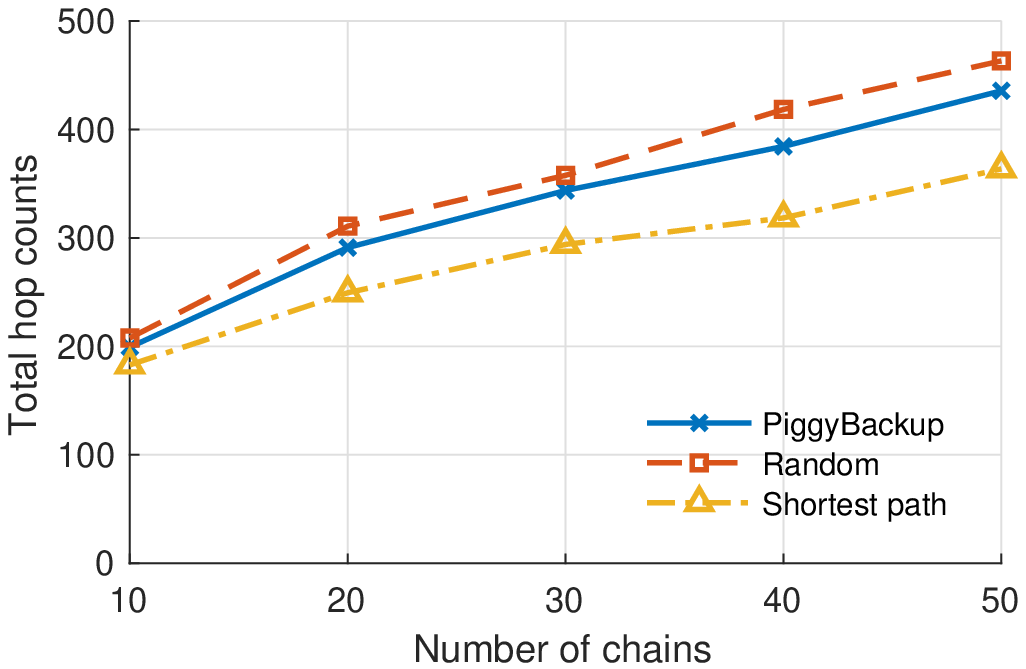}\\
				(b) total number of hop counts\\
				\epsfig{width=2.8in,file=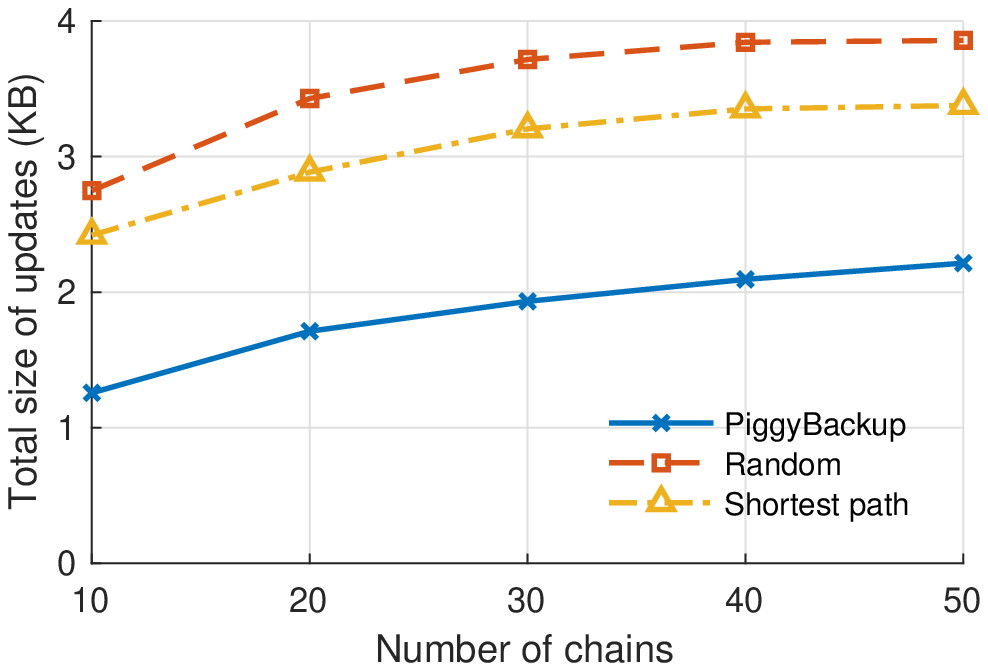}\\
				(c) total size of updates
			\end{tabular}
			\caption{Impact of number of chains.}
			\label{fig:chainNum}
		\end{figure}
		}
		\vspace*{\fill}
		%\clearpage
	}
\fi

\secc{Performance Evaluation}\label{sec:results}
We conduct extensive simulations written in C to evaluate the
performance of \name\ in a fat tree topology~\cite{AlFares2008}, which
has been widely used in many data center networks. We deploy a 4-pods
fat tree, which consists of 16 hosts and three layers of switches,
i.e., 8 edge switches, 8 aggregation switches and 4 core switches.
Each switch connects to a physical machine, which installs a number of
VMs.  Without otherwise stated, the default number of function types
and chains are set to 20 and 50, respectively, and the maximal number
of primary instances and backup instances that can be installed in a
server (i.e., server capacity) is 8 and 3, respectively.  We assign
each primary instance to a randomly-selected physical machine. For
each function type, we randomly deploy a roughly equal number of
instances in the network until the capacity of all the servers has
been fully utilized. Each chain requesting for a function type is
assigned to the closest function instance. That is, for a chain
requesting the functions $f_i$ and $f_j$ sequentially, we assign the
chain to the instance of $f_j$ closest to the server of $f_i$'s
instance.  Each backup instance can be associated with up to $K=5$
primary instances by default.  The packet size of each piggyback
update and stand-alone update, respectively, is set to 20 and 60
bytes.  Each simulation repeats for 20 random rounds, and we report
the average performance. The default parameter settings are summarized
in Table~\ref{tb:setting}.
%\del{\textblue{The default parameter settings are summarized in

We generate a set of chains, each of which is from a random
originating host to a random destination host and requests a sequence
of NFs randomly selected from a set of function types $\mathcal{F}$.
The length of a chain is also picked randomly, varying from 1 to 20.
We leverage a simple closest instance assignment scheme to assign
primary instances to a chain. That is, for a chain requesting NF
$f_i\rightarrow f_{i+1}$ in steps $i$ and $i+1$, the instance of $f_i$
will forward this chain to the closest instance of $f_{i+1}$. The
traffic arrival pattern of each chain follows a Poisson process with
the mean arrival rate of 1 packet per millisecond. Since our goal is
to demonstrate the benefit of reusing SFCs, the primary instance
assignment for SFCs should not affect the performance much.

We compare \name\ with two baseline algorithms: {\em i) random
placement}, which deploys backup instances in randomly selected
servers without considering the locations of primary instances, and
{\em ii) shortest path based placement:} primary instances
sequentially pick the closest available servers to deploy (or reuse)
their backup instances, while the selection order of primary instances
is determined randomly.  Our evaluation will examine the impact of
number of chains, the impact of server capacity, the impact of
backup-primary association limit, the effectiveness of chain
selection, and the performance of stand-alone backup instance
placement.

\ifx\isthesis\undefined
\begin{figure}[t]
	\centering
	\begin{tabular}{ccc}
		%\hspace{-12pt}
		\epsfig{width=2.8in,file=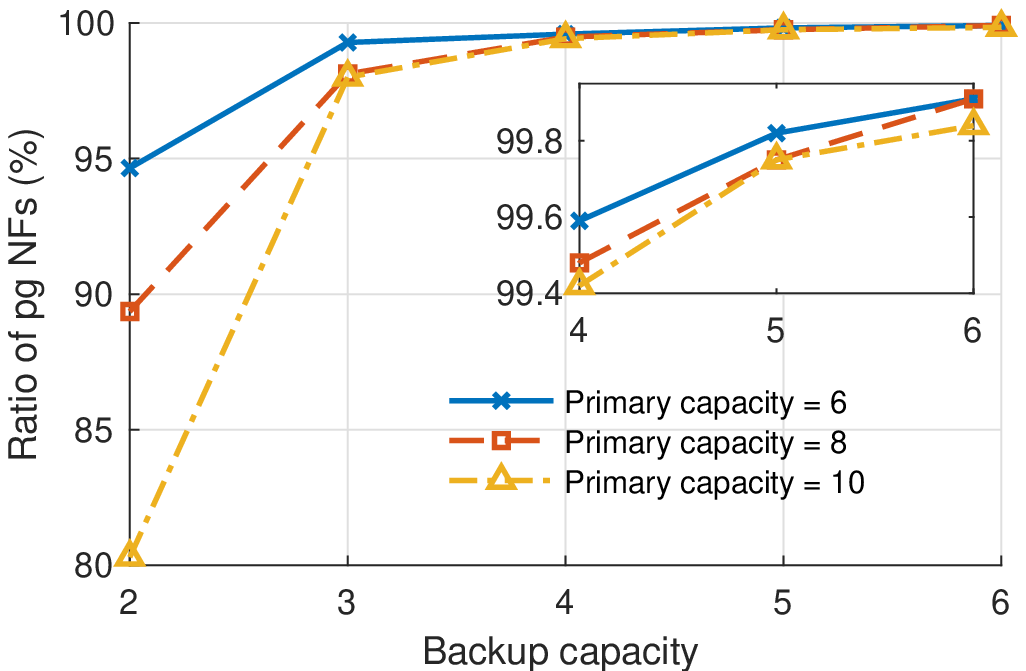}\\
		\vspace{6pt}
		\footnotesize{(a) 10 chains}\\
		%\hspace{-10pt}
		\epsfig{width=2.8in,file=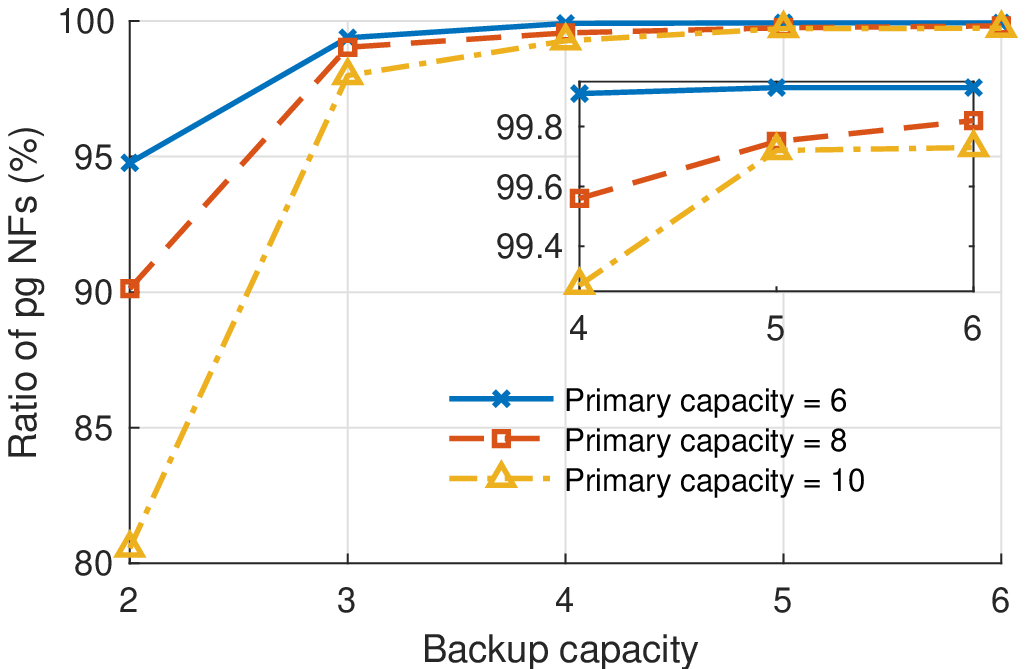}\\
		\vspace{6pt}
		\footnotesize{(b) 20 chains }\\
		%\hspace{-10pt}
		\epsfig{width=2.8in,file=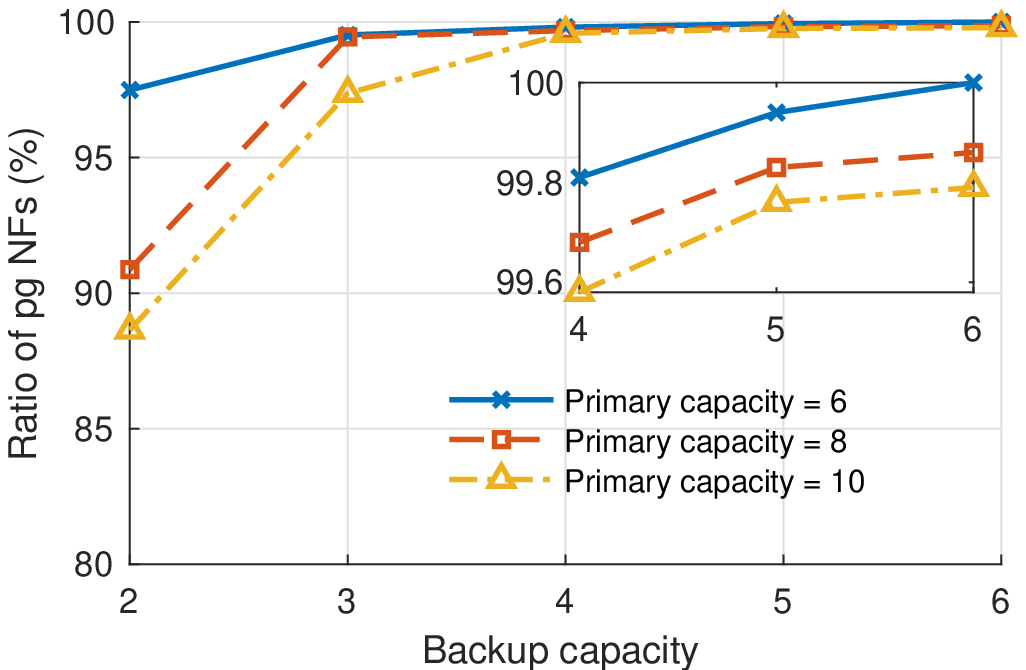}\\
		\footnotesize{(c) 30 chains }
	\end{tabular}
	\caption{Impact of backup capacity}
	\label{fig:capacity}
\end{figure}
\else
%\afterpage{
	\vspace*{\fill}{
		\begin{figure}[ht]
			\centering
			\begin{tabular}{c}
				\epsfig{width=4in,file=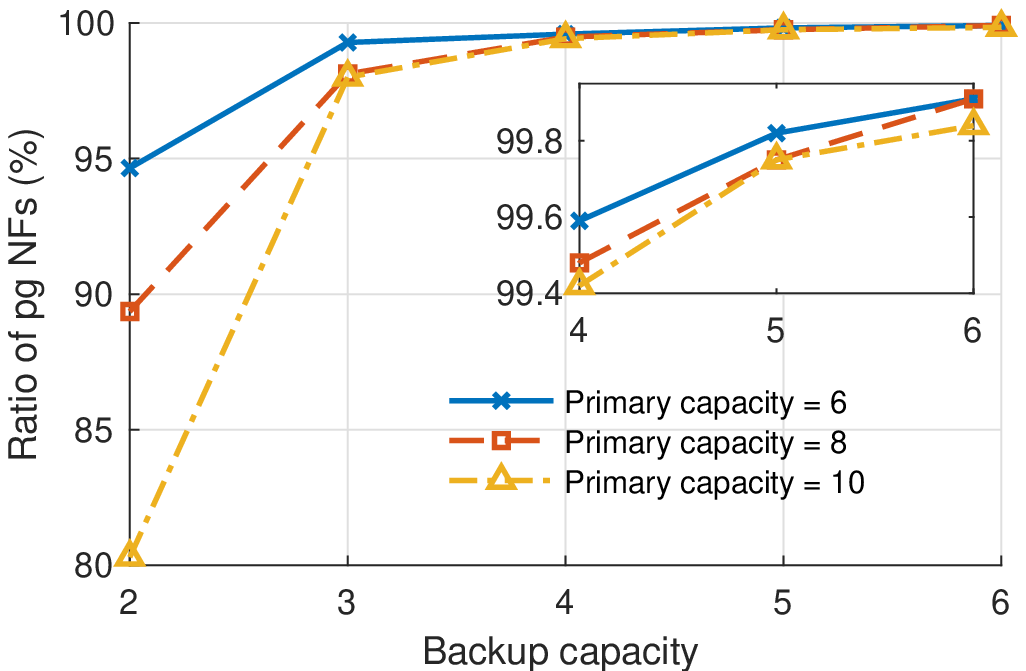}\\
				\footnotesize{(a) 10 chains}\\
				\\
				\epsfig{width=4in,file=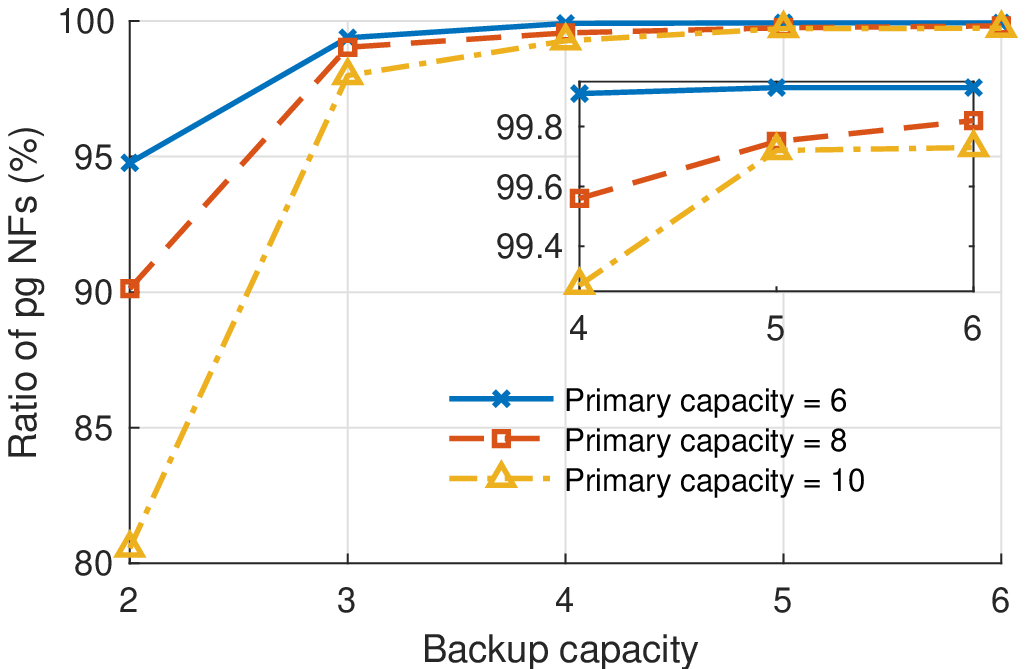}\\
				\footnotesize{(b) 20 chains }\\
				\\
				\epsfig{width=4in,file=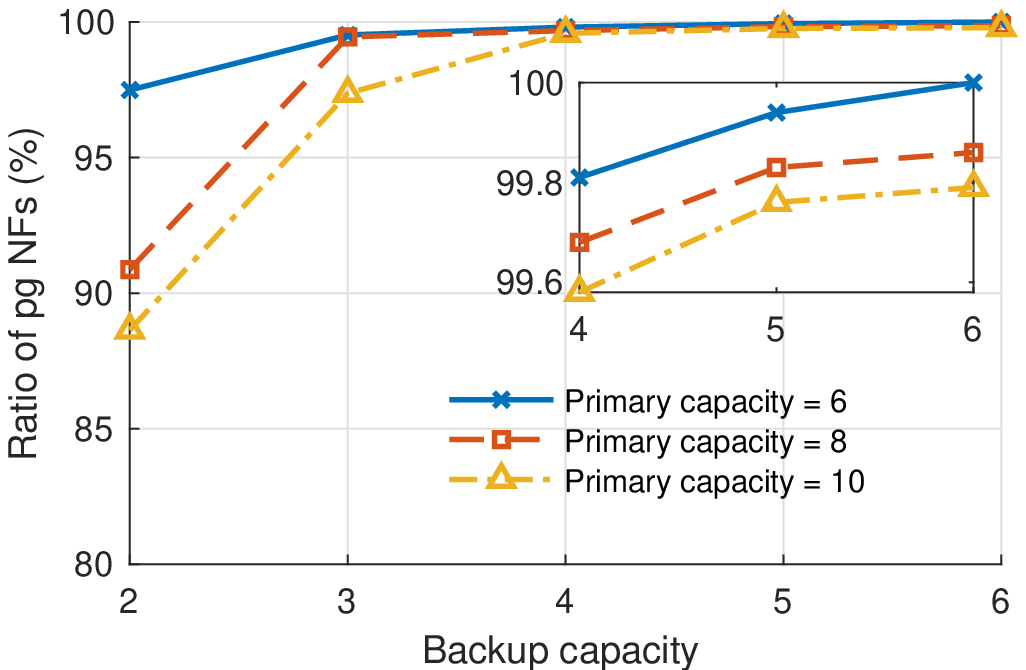}\\
				\footnotesize{(c) 30 chains }
			\end{tabular}
			\caption{Impact of backup capacity}
			\label{fig:capacity}
		\end{figure}
	}
	\vspace*{\fill}
	%\clearpage
%}
\fi

\ifx\isthesis\undefined
        \begin{figure}[t]
	        \centering
		    \epsfig{width=2.8in,file=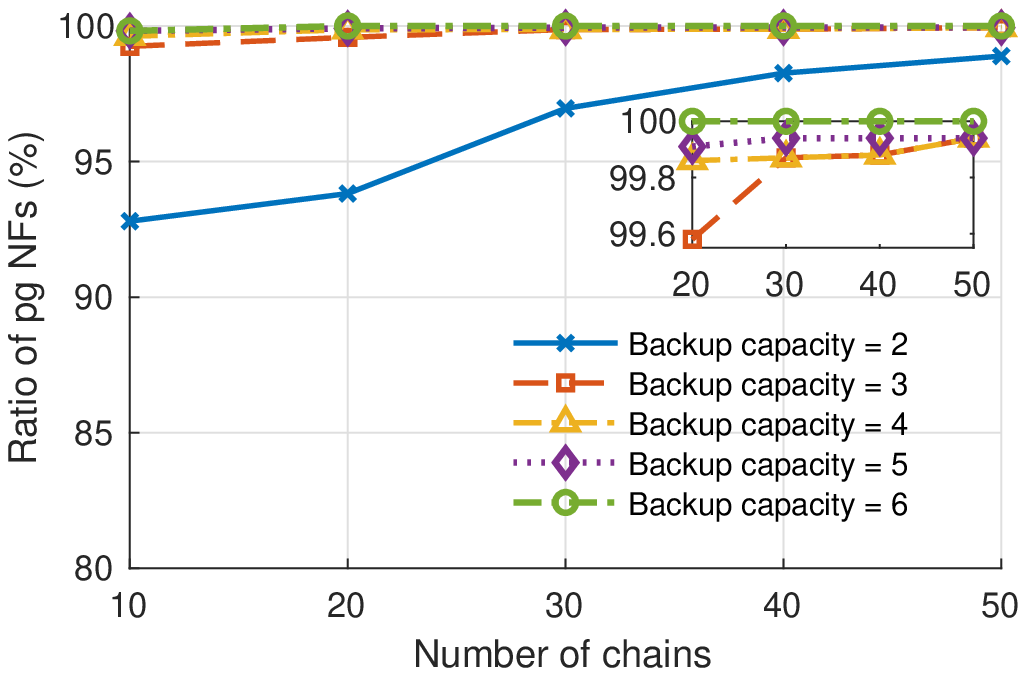}
	        \caption{Impact of number of chains and backup capacity}
	        \label{fig:chain-cap}
        \end{figure}
\else
\fi

\subsecc{Impact of Number of Chains}%Fig 4,5,6
We first check the percentage of primary instances whose backup can be
piggybacked in the three comparison schemes.
Fig.~\ref{fig:chainNum}(a) shows the results when the number of chains
varies from 10 to 50. The figure shows that, as the number of chains
increases, the opportunities of piggybacking grow accordingly,
matching the intuition of our design.  Since our design explicitly
considers the locations of primary instances and the traversing paths
of chains, we can fully utilize the piggybacking opportunities and
obtain almost 100\% of piggybackable instances. However, random
placement and shortest path based deployment would underutilize the
opportunities as they do not consider the routing of chains.

Fig. \ref{fig:chainNum}(b) demonstrates the piggybacking hop counts of
the comparison methods. The results show that random placement does
not consider the locations of primary instances and, thus, lead to long
piggybacking paths.  Shortest path based placement achieves the minimal
piggybacking hop counts since it always deploys the backup instances
nearby the primary instances. However, as shown in
Fig.~\ref{fig:chainNum}(a), it supports piggybacking for only
40\%--60\% of primary instances. Hence, those non-piggybackable
instances still need to use stand-alone packets to deliver updates, as
a result increasing the overall cost. By contrast, \name\ allows
almost all the primary instances to piggyback their updates, through a
slightly longer path.

We then plot in Fig.~\ref{fig:chainNum}(c) the total number of update
bytes, including piggybacking and stand-alone packets, of the three
methods for every update event. The results verify that shortest path
based placement sends stand-alone updates for a large proportion of
primary instances and, overall, produces a significantly higher update
overhead.  Though our piggybacking updates traverse through slightly
longer paths, the average overall overhead can still be reduced by
39.56\%, as compared to shortest path based placement, since we fully
utilize piggybacking opportunities and minimize the stand-alone update
cost. While the total size of every update event is only a few KBs,
the overhead saving can be significant when there are many functions
coexisting in the network and updating frequently at the same time.

\subsecc{Impact of Server Capacity}%Fig 4,5,6
In this simulation, we examine the impact of backup capacity of each
server.  Figs.~\ref{fig:capacity}(a,b,c) illustrate the percentage of
piggybackable instances as the network includes 10, 20, and 30 chains,
respectively. For each figure, we plot the results when the backup
capacity varies from 2 to 6.  The results show that, when the backup
capacity increases, the piggybacking opportunities grow accordingly.
The percentage converges when the backup capacity increases to 4,
showing that a finite backup capacity can already ensure full
piggybacking updates. When the backup capacity is large enough, the
performance of various numbers of chains becomes indifferent. This
explains that, when a system supports sufficient backup capacity, even
a small number of chains can already realize the benefits of
piggybacking.

\ifx\isthesis\undefined
        \begin{figure}[t]
	        \centering
		    \epsfig{width=2.8in,file=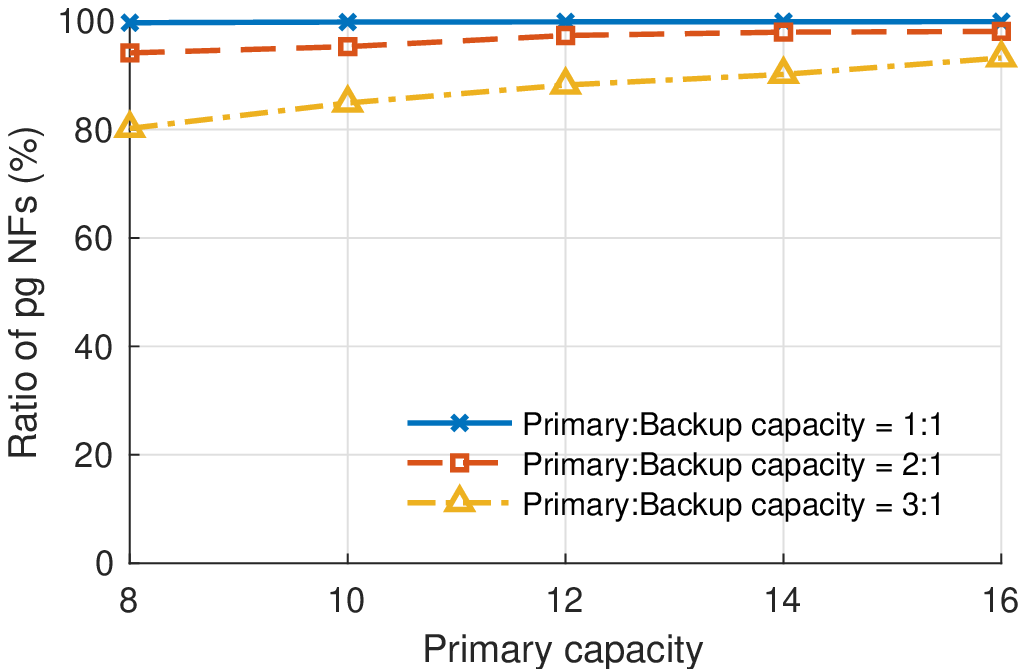}
	        \caption{Impact of primary-to-backup capacity ratio}
	        \label{fig:p-bRatio}
        \end{figure}
\else
\afterpage{
    \vspace*{\fill}{
        \begin{figure}[h]
	        \vspace{-2mm}
	        \centering
		    \epsfig{width=4in,file=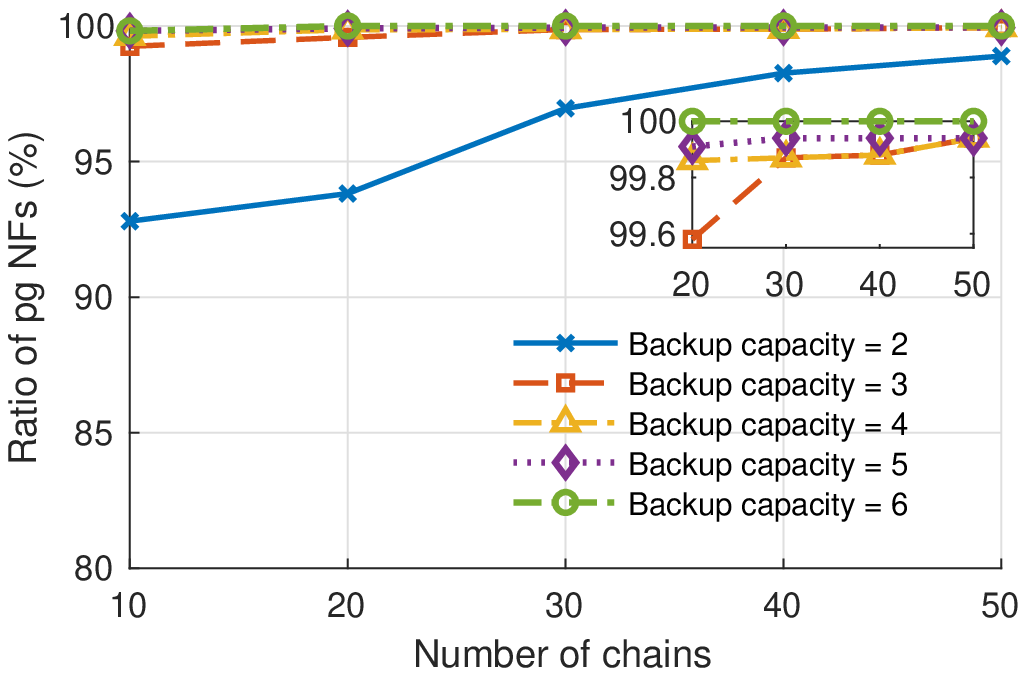}
	        \caption{Impact of number of chains and backup capacity}
	        \label{fig:chain-cap}
        \end{figure}
        \begin{figure}[h]
	        \vspace{-2mm}
	        \centering
		    \epsfig{width=4in,file=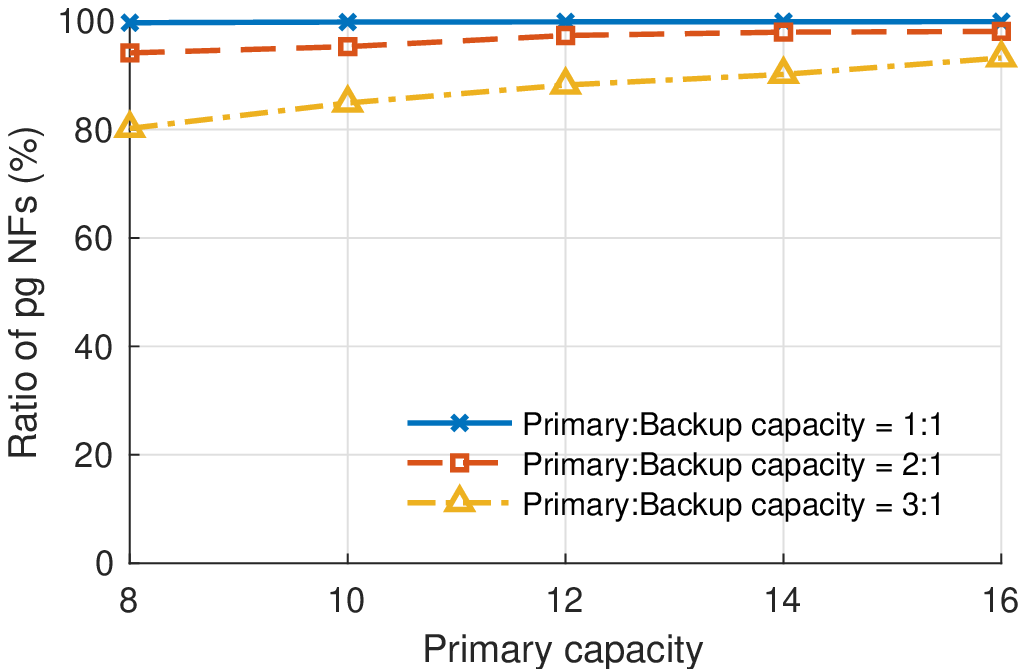}
	        \caption{Impact of primary-to-backup capacity
			ratio}
	        \label{fig:p-bRatio}
        \end{figure}
    }
    \vspace*{\fill}
    \clearpage
}
\fi

Fig.~\ref{fig:chain-cap} summarizes the piggybacking percentage for
various numbers of chains and backup capacity. The primary capacity of
each server is fixed to 8. The figure shows a similar trend, in which
the piggybacking opportunities can be mostly utilized when the backup
capacity grows above 3, regardless off the number of chains in the
network.  Even if each server can only install two backup instances,
\name\ can still allow more than 93\% of instances to piggyback their
updates.  We next fix the total capacity of each server, but change
the ratio of resources allocated to primary and backup instances.  For
example, a 3:1 ratio implies that, in a server, 75\% of instances are
primary instances, while 25\% are backup instances.
Fig.~\ref{fig:p-bRatio} shows the percentage of piggybackable
instances as the primary capacity varies from 8 to 16.  The results
demonstrate that the piggybacking opportunities are closely related to
how much percentage of server resources is used for backup. If servers
reserve fewer resources for backup, the piggybacking opportunities
degrade. Resource allocation for primary and backup instance
deployment is an interesting issue worth future studies.

\ifx\isthesis\undefined
\begin{figure}[t!]
	\centering
	\epsfig{width=2.8in,file=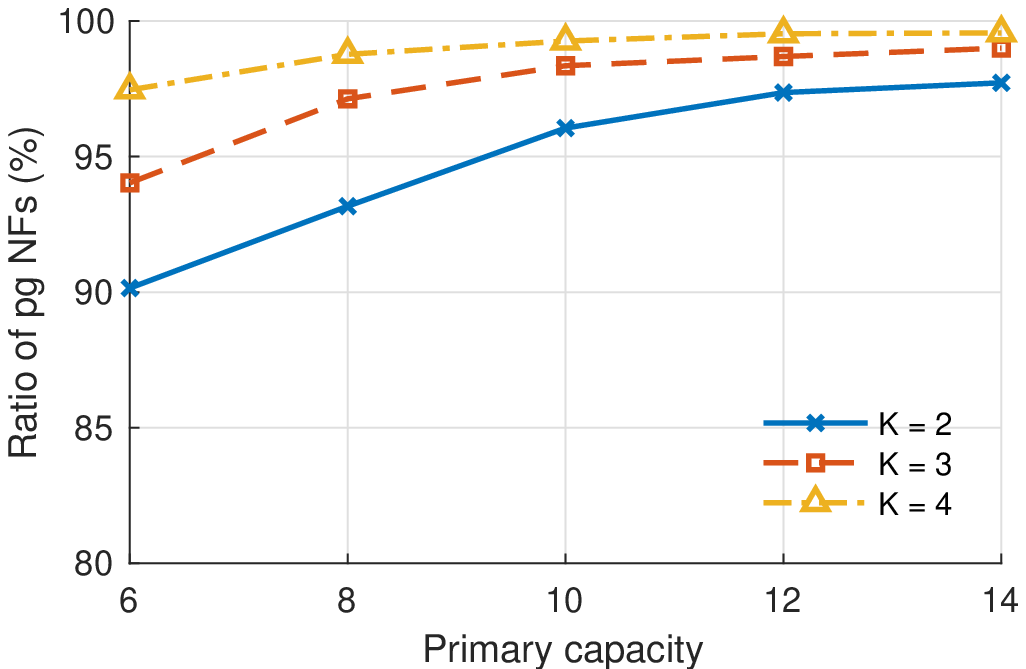}
	\caption{Impact of backup-primary association limit (P:B capacity =
	2:1)}
	\label{fig:impactK}
\end{figure}

\begin{figure}[t!]
	\centering
	\epsfig{width=2.8in,file=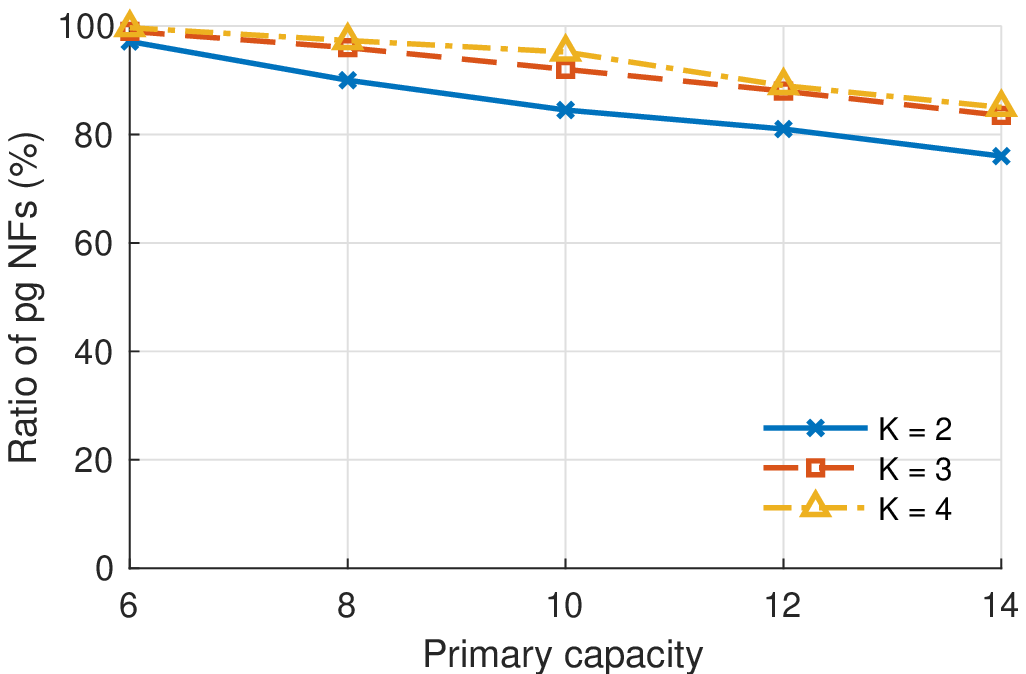}
	\caption{Impact of backup-primary association limit (Backup capacity
	= 3)}
	\label{fig:impactK2}
\end{figure}
\else
\afterpage{
	\vspace*{\fill}{
		\begin{figure}[h!]
			\centering
			\epsfig{width=4in,file=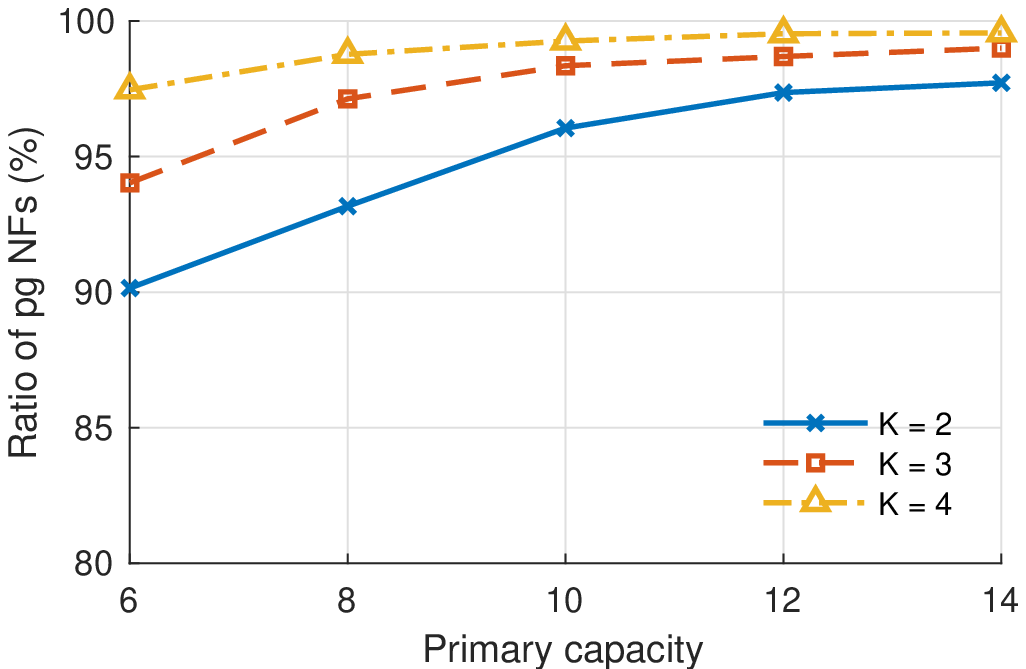}
			\caption{Impact of backup-primary association limit
			(Primary:Backup capacity = 2:1)}
			\label{fig:impactK}
		\end{figure}

		\begin{figure}[h!]
			\centering
			\epsfig{width=4in,file=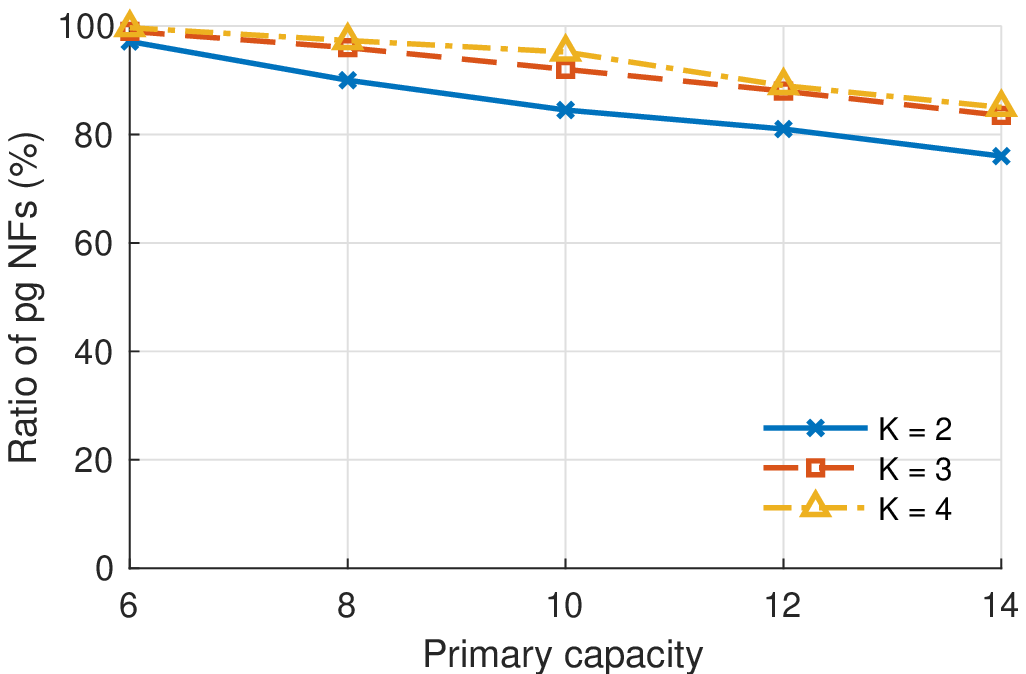}
			\caption{Impact of  backup-primary association limit (Backup
			capacity = 3)}
			\label{fig:impactK2}
		\end{figure}
		}
		\vspace*{\fill}
		\clearpage
	}
	\fi

\subsecc{Impact of Backup-Primary Association Limit}%Fig 11
We then check how the performance changes with the value of $K$
(maximal number of primary instances that can be associated with a
backup instance).  We try two different configurations. In
Fig.~\ref{fig:impactK}, we fix the ratio of primary capacity to backup
capacity per server to 2:1, while in Fig.~\ref{fig:impactK2}, we fix
the back capacity to 3 but vary the primary capacity from 6 to 14.
Recall that a smaller $K$ means that each backup instance can only
serve fewer primary instances. That is, the degree of competition
among primary instances becomes more severe when the system is short
of backup instances. The figures confirm again that the piggybacking
opportunities decrease as $K$ gets smaller.

Fig.~\ref{fig:impactK} shows that, when the overall capacity per
server is small, e.g., 6 primary instances per server, only a few
backup instances are deployed and the piggybacking opportunities
become small since fewer chains can traverse through a limited number
of backup instances. Similarly, given a fixed number of backup
instances, as in Fig.~\ref{fig:impactK2}, an increasing number of
primary instances leads to more severe competition, as a result making
some primary instances not be able to find any piggybackable chains.
The results reveal that piggybacking opportunities not only depend on
the number of chains but also the number of back instances distributed
in the network.

\ifx\isthesis\undefined
        \begin{figure}[t]
	        \centering
		    \epsfig{width=2.8in,file=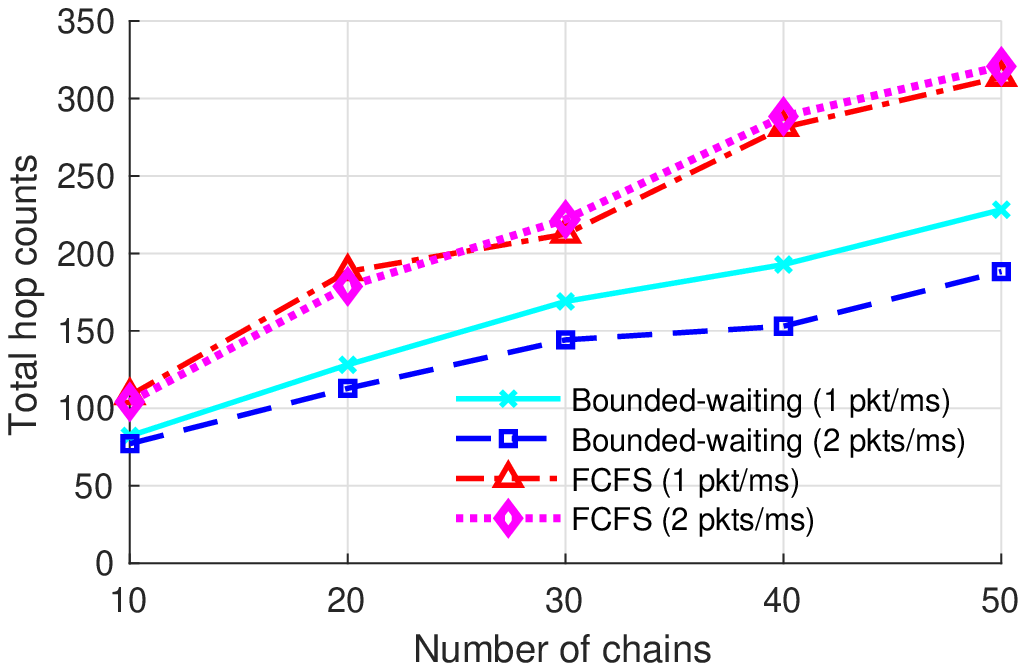}
	        \caption{Effectiveness of chain selection}
	        \label{fig:chainselection}
        \end{figure}
\begin{figure}[t!]
\centering
\epsfig{width=2.8in,file= 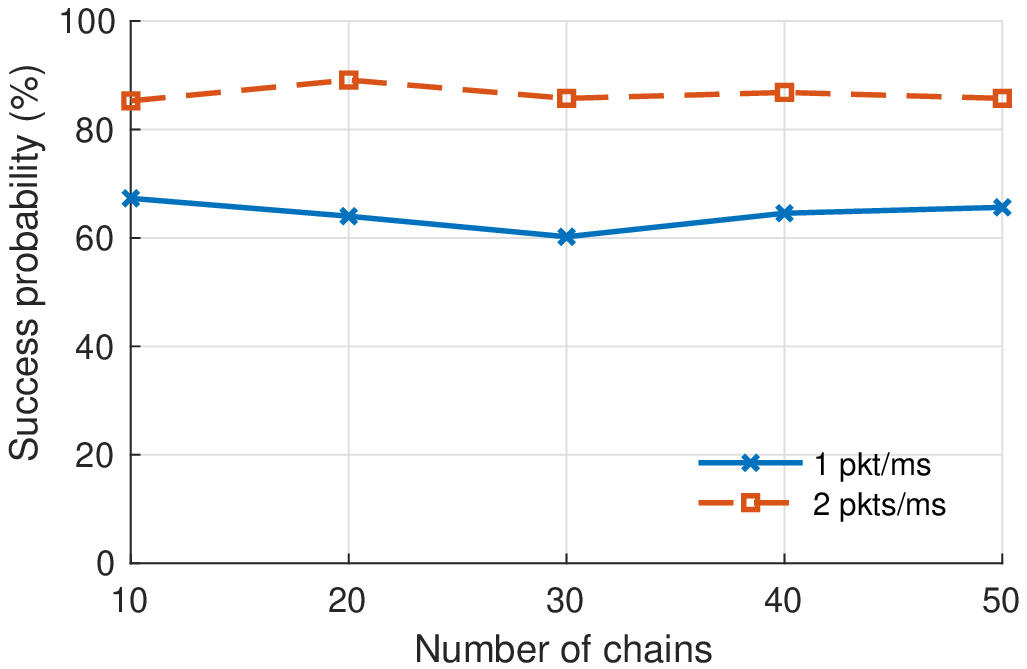}
\caption{Success probability of chain selection}
\label{fig:successprob}
\end{figure}
\else

\afterpage{
    \vspace*{\fill}{
        \begin{figure}[h]
	        \centering
		    \epsfig{width=4in,file=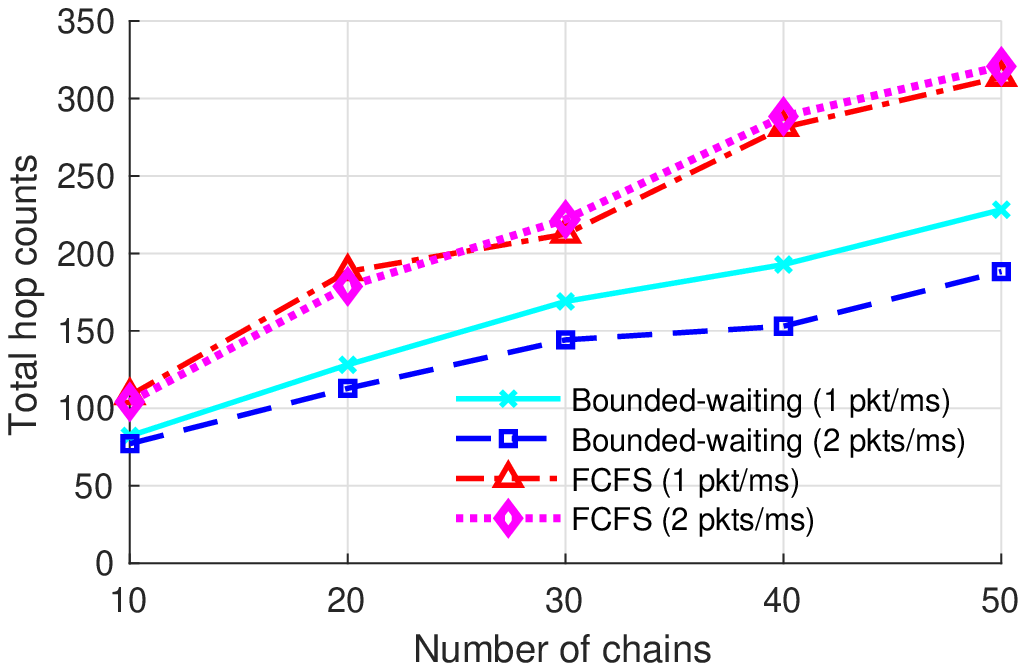}
	        \caption{Effectiveness of chain selection}
	        \label{fig:chainselection}
        \end{figure}
        \begin{figure}[h]
	        \centering
		    \epsfig{width=4in,file=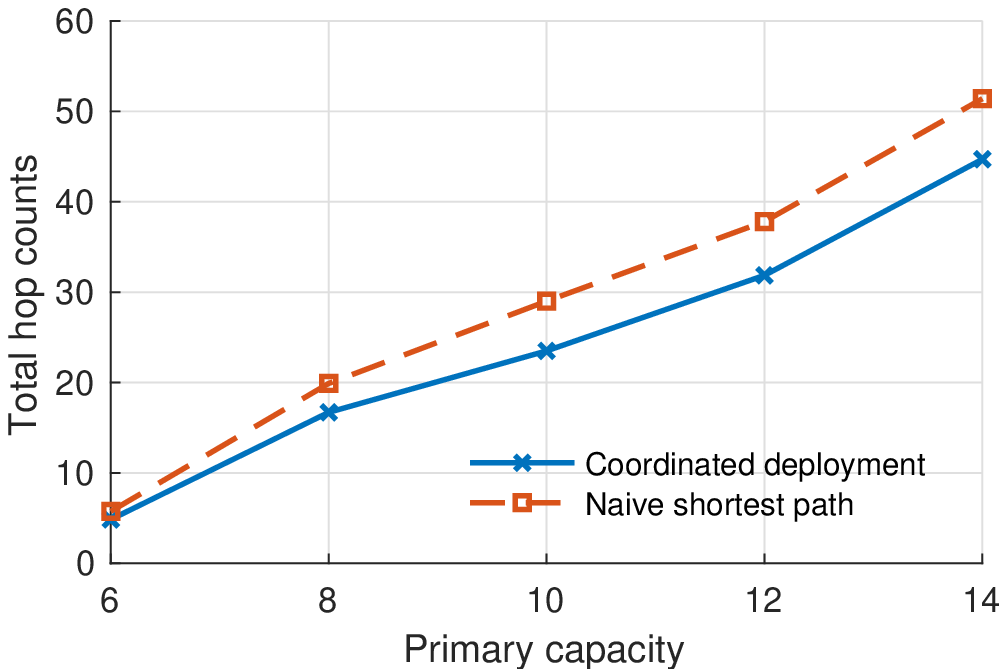}
	        \caption{Performance of stand-alone instance deployment}
	        \label{fig:stand-alone}
        \end{figure}
    }
    \vspace*{\fill}
    \clearpage
}
\fi

\subsecc{Effectiveness of Chain Selection}%Fig 11
We next check whether our chain selection design can effectively
reduce the piggybacking cost as the mean packet arrival rate is
configured to 1 and 2 packets per millisecond, respectively.  We also
compare our chain selection with na\"{\i}ve FCFS. The results
illustrated in Fig.~\ref{fig:chainselection} show that FCFS introduces
a much higher piggybacking hop count since primary functions are more
likely to piggyback the update states in a chain traversing along a
long path.  With our arrival prediction, we try to embed the updates
in shorter chains if possible, as a result reducing the hop count by
around 27.5\% and 39.35\% with respect to the packet arrival rate of 1
and 2 packets per millisecond, respectively. The figure also shows
that a higher traffic rate increases the probability of identifying a
short piggybacking chain and results in a smaller piggybacking cost.

We further plot the success probability of chain selection in
Fig.~\ref{fig:successprob} when the packet arrival rate is configured
to 1 and 2 packets per millisecond, respectively. The success
probability here means the number of epochs whose candidate identified
by our chain selection actually arrives in that epoch divided by the
total number of epoches. For example, if our candidate selection in 80
out of 100 epochs can actually arrive in the epoch and help piggyback
the update, we say that the success probability equals 80\%. in The
results show that, when the packet arrival rate is sufficiently high
(2 packets per ms), over 80\% of our prediction can correctly identify
the arrival events of the candidate chains, as a result helping
piggyback the update through a shorter path and reducing the
forwarding latency effectively. Even for a lower arrival rate (1
packet per ms), we can still identify the candidates with a
probability of around 60\%--70\%.  The lower probability is due to the
few arrival events. Overall, we can conclude that, with arrival
prediction, each primary function can have a fairly high opportunity
to piggyback the update information in the identified shorter chain
candidate and, thus, reduce the update cost.

\begin{figure}[t]
	\centering
	\begin{tabular}{c}
		%\hspace{-12pt}
		\epsfig{width=2.8in,file=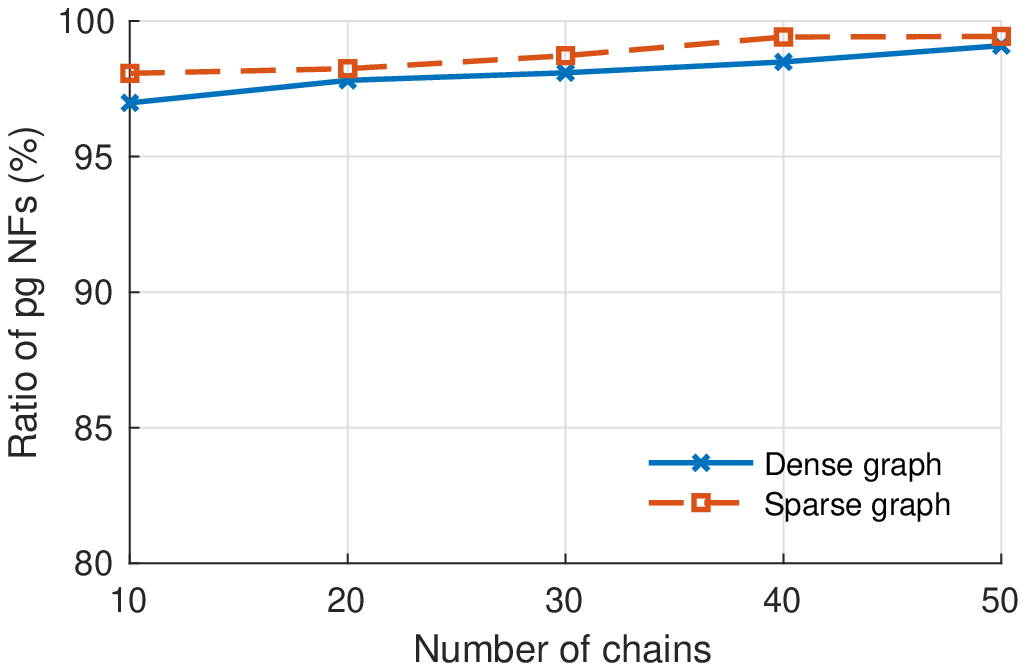}\\
		\vspace{6pt}
		\footnotesize{(a) percentage of piggybackable instances}\\
		%\hspace{-10pt}
		\epsfig{width=2.8in,file=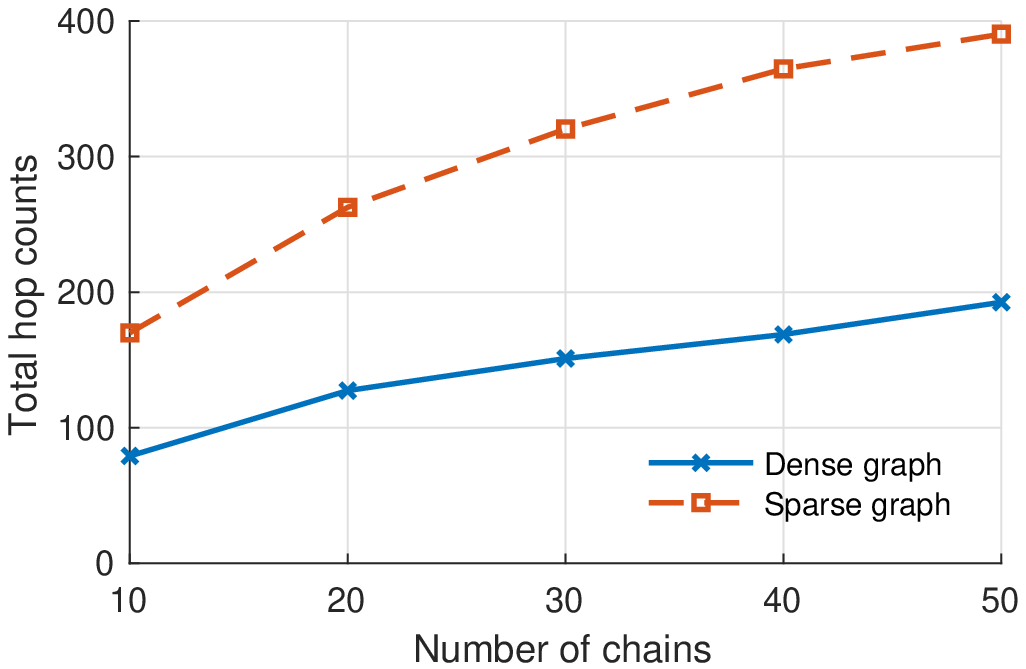}\\
		\vspace{6pt}
		\footnotesize{(b) total number of hop counts }\\
		%\hspace{-10pt}
		\epsfig{width=2.8in,file=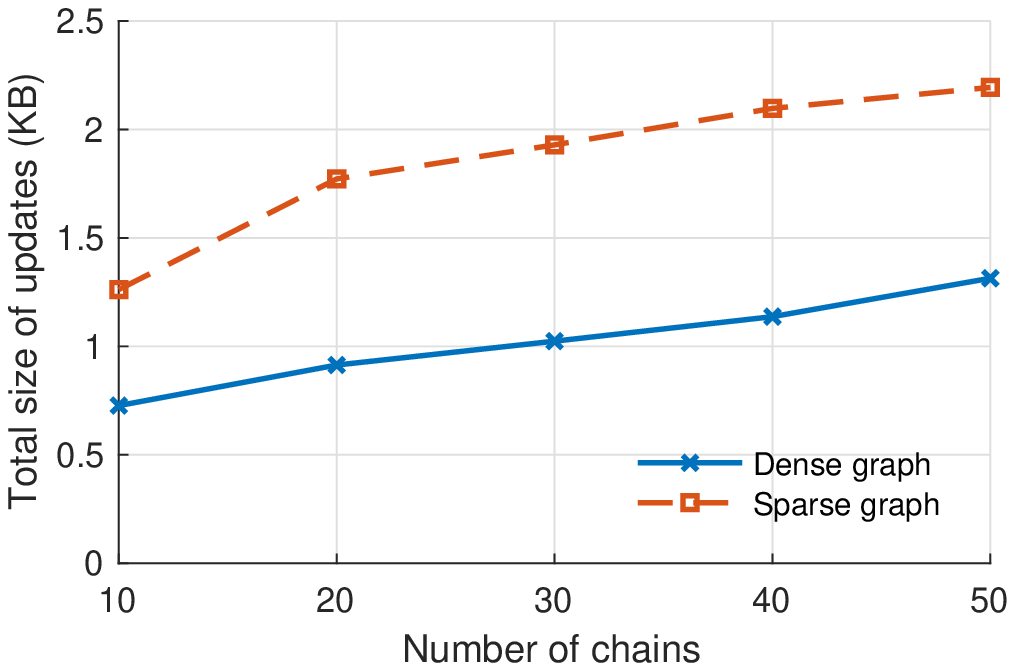}\\
		\footnotesize{(c) total size of updates}
	\end{tabular}
	\caption{Impact of network density}
	\label{fig:topology}
\end{figure}

\subsecc{Impact of Network Density}
We next check the impact of network topology on the update costs.  In
this simulation, we deploy 20 servers but connect any two servers with
a random probability, following the uniform distribution.  To tune the
density of a network, we set the mean connection probability to 20\%
and 80\% to simulates a sparse and dense network, respectively.
Figs.~\ref{fig:topology}(a), ~\ref{fig:topology}(b) and
\ref{fig:topology}(c) illustrates the ratio of piggybackable
instances, the total hop counts of piggybacking and the total update
costs, respectively, of the two different topologies.
Fig.~\ref{fig:topology}(a) shows that the piggyback ratio of a sparse
graph is higher than that in a dense graph. The reason is that, when a
network is sparse, SFCs only have a few available paths and need to go
through more network links, which hence increases the probability of
traversing through multiple primary instances and backup instances.
This can also be observed in Fig.~\ref{fig:topology}(b), which
illustrates that the piggybacking path in a sparse graph is far longer
than that in a dense graph.  Finally, Fig.~\ref{fig:topology}(c)
indicates that, though a dense graph has fewer piggybacking
opportunities, it has more links and usually results in a shorter
piggybacking path. Hence, the overall update cost of \name\ in a dense
graph is much lower than that in a sparse graph.

        \begin{figure}[t]
	        \centering
		    \epsfig{width=2.8in,file=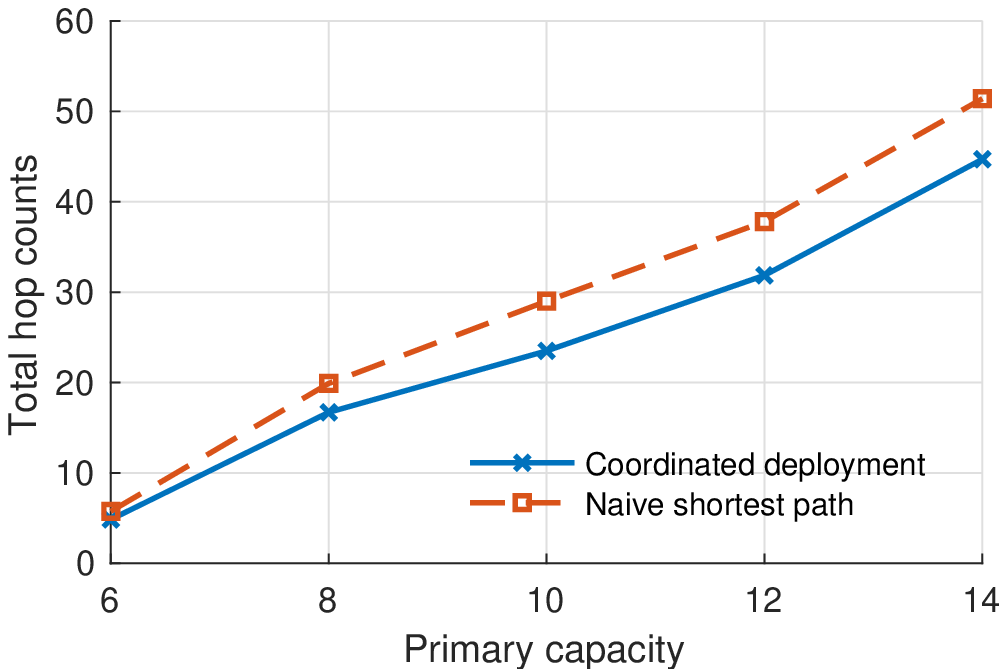}
	        \caption{Performance of stand-alone instance deployment}
	        \label{fig:stand-alone}
        \end{figure}

\subsecc{Performance of Stand-Alone Backup}%Fig 11
Finally, Fig.~\ref{fig:stand-alone} compares the backup hop count of
\name's cooperative stand-alone instance deployment and na\"{\i}ve
shortest path based deployment. The results show that the stand-alone
cost increases as the number of primary instances per server increases
due to a limited number of chains and, thus, limited piggybacking
opportunities.  By explicitly controlling the backup placement order
of prioritized primary instances, we can better utilize the remaining
available resources to reduce the cost of non-piggybackable primary
instances, as compared to the simple shortest path based scheme. The
gap between the two solutions grows as a network has more primary
instances, which should compete for limited backup capacity. Recall
the results shown in Fig.~\ref{fig:chainNum}(c). The reduced overall
update cost in terms of update bytes is also contributed by a more
efficient cooperative stand-alone instance deployment scheme of \name.

\secc{Conclusion and Future Work}\label{sec:conclusion}
In this \article, we presented a backup NF instance placement
framework to provide fault tolerance and ensure state consistency for
a software-defined network.  Our design reuses existing service chains
to {\em piggyback} backup information and, thereby, reduces network
usage for periodical state updates.  We have derived the joint backup
deployment and assignment problem as an ILP model and proposed
heuristic algorithms to enhance piggybacking opportunities and
minimize the stand-alone update cost.  A chain selection strategy was
also developed to identify shorter chains for piggybacking subject to
a constrained update latency.  The simulation results show that the
more SFCs available in a network, the higher gain \name\ can achieve,
verifying the benefits of piggybacking.  Overall, \name\ reduces the
update traffic load by 39.56\% and 47.65\%, respectively, on average
in a fat-tree topology as compared to shortest path based placement
and random placement. 

\ifx\isthesis\undefined
%~
\else
Some future directions are worth further studied.  First, we now
associate each primary function instance with only a single backup
instance. The failure could still occur when both instances fail
simultaneously. We can investigate how to associate each primary
instance with multiple backup instances so as to ensure a predefine
reliability level. Second, different physical machines and network
functions would have heterogeneous failure probability. While our work
now considers homogeneous failure probability, we could extend our
design to further consider heterogeneous failure probability.  Third,
the placement of primary instances and backup instances can be jointly
considered to optimize piggybacking opportunities and minimize the
update costs. Finally, while our work reuses existing SFCs, whose NF
assignment and routing have been determined, we can also flexibly
adapt the assignment for SFCs to enhance piggybacking opportunities
while maintaining their quality of service.

\begin{comment}
in general network topology, multiple backup instances for one primary
function instance have potential to further enhance fault tolerant,
because one to one mapping of primary instances and backup instances
has higher chance that the primary instance and its corresponding
backup instance encounter failures simultaneously.  Second, different
physical machines will have different failure probability, our model
and algorithms can also take this factor into consideration to
optimize the reliability of the whole network. Third, we can jointly
consider primary instances and backup instances deployment, different
primary instance placement may make chains go through different path,
and affect if chains can help piggyback update information or the
length of piggybackable traffic. Forth, we can assign instances for
SFCs with consideration of piggybacking opportunities to achieve high
percentage of piggybackable functions.
\end{comment}
\fi

%\del{\textblue{write more about future work}

\bibliographystyle{IEEEtran}
\bibliography{paper}
\end{document}